\documentclass[prl,twocolumn,showpacs,floatfix]{revtex4}
\usepackage{times}

\usepackage{graphicx}
  \graphicspath{{./Eps/}}
\usepackage{amsmath}
\usepackage{amssymb}

\newcommand{\vc}[1]{
\textbf{\textit{#1}} }

\newcommand{\equr}[2]{
\begin{eqnarray}
#2
\label{#1}
\end{eqnarray}}

\newcommand{\dif}{\partial}

\newcommand{\av}[1]{\langle #1 \rangle}

\newcommand{\schre}{Schr\"{o}dinger }

\begin{document}

\title{Conductance and persistent current in quasi-one-dimensional systems with grain boundaries: Effects of the strongly reflecting and columnar grains}

\author{J. \surname{Feilhauer}}

\author{M. \surname{Mo\v{s}ko}}
\email{martin.mosko@savba.sk}

\affiliation{Institute of Electrical Engineering, Slovak
Academy of Sciences, 841 04 Bratislava, Slovakia}

\date{\today}

\begin{abstract}
We study mesoscopic transport in the quasi-one-dimensional wires and rings
made of a two-dimensional conductor of width $W$ and length $L \gg W$. Our aim is to compare an impurity-free conductor with grain boundaries with a grain-free conductor with impurity disorder.  A single grain boundary is modeled as a set of the two-dimensional-$\delta$-function-like barriers positioned equidistantly on a straight line and disorder is emulated by a large number of such straight lines, intersecting the conductor with random orientation in random positions. The impurity
disorder is modeled by the two-dimensional
$\delta$-barriers with the randomly chosen positions
and signs.
The electron transmission through the wires is calculated by
the scattering-matrix method, and the Landauer conductance is obtained.
Moreover, we calculate the persistent current in the rings threaded by magnetic flux: We incorporate into the scattering-matrix method the
flux-dependent cyclic boundary conditions and we introduce a trick allowing to study the persistent currents in rings of almost realistic size.
We mainly focus on the numerical results for $L$ much larger than the electron mean-free path, when the transport is diffusive. If the grain boundaries are weakly reflecting, the systems with grain boundaries show the same (mean) conductance and the same (typical) persistent current as the systems with impurities, and the results also agree with the single-particle theories treating disorder as a white-noise-like potential. If the grain boundaries are strongly reflecting, the rings with the grain boundaries show the typical persistent currents which can be about three times larger than the results of the white-noise-based theory, thus resembling the experimental results of Jariwala et al., Phys. Rev. Lett. \textbf{86}, 1594 (2001). Finally, we extend our study to the three-dimensional conductors with columnar grains. We find that the persistent current exceeds the white-noise-based result by another one order of magnitude, similarly as in the experiment of Chandrasekhar et al., Phys. Rev. Lett. \textbf{67}, 3578 (1991).
\end{abstract}


\maketitle


\section{I. Introduction}

Magnetic flux $\Phi$ piercing the opening of a mesoscopic conducting ring gives rise to the equilibrium electron current circulating along the ring. This current is known as persistent current \cite{Imry-book}. A single electron at the energy level $E_n$ carries the current $I_n = -\partial E_n(\Phi)/\partial \Phi$ \cite{buttiker83,byers}. At zero temperature the persistent current in the ring is given as $I = \sum I_n$, where one sums over all occupied states below the Fermi level \cite{Cheung1D}. In a single-channel ballistic ring, the amplitude of the persistent current is $I_0 = ev_F/L$, where $e$ is the electron charge, $v_F$ is the electron Fermi velocity and $L$ is the ring circumference. The current changes its sign when a single electron is added into the ring. In the ballistic ring with $N_c$ conducting channels the amplitude of the persistent current scales as $\sqrt{N_c}I_0$ due to the random sign of the current in each channel \cite{IBM}.

If disorder is present,
the persistent current does not vanish and the amplitude and sign depend on the specific configuration of disorder. Therefore, it is customary to study the typical persistent current $I_{typ} = \langle I^2 \rangle^{1/2}$, where $\langle \dots \rangle$ is the ensemble average. The authors of the works \cite{Cheung,riedel} analyzed the persistent current in a disordered normal-metal multi-channel ring by considering the non-interacting electron gas.
They assumed that the electrons are scattered by disorder with the scattering potential $V(\bold{r})$ obeying the white-noise condition $\langle V(\bold{r}) V(\bold{r}') \rangle \propto \delta(\bold{r} - \bold{r}')$. Using the Green's functions, they
found for the typical current at $\Phi = \pm 0.25 h/e$ the formula
\begin{equation}
I^{theor}_{typ} = 2 \times \frac{1.6}{d} I_0 \frac{l}{L}, \quad l \ll L \ll \xi,
 \label{Igre}
\end{equation}
where the factor of 2 is due to the spin, $l$ is the electron mean free path, $\xi$ is the localization length, $d$ is the dimensionality of the ring, and
the condition $l \ll L \ll \xi$ means the diffusive regime. (The factor $1.6/d$ is derived in the appendix A from a more general formula from the literature.)

The persistent current in a single isolated ring was for the first time measured by Chandrasekhar \textit{et al.} \cite{Chand}. In that experiment, three different Au rings of size $L \sim 100 l$ showed the persistent currents ranging from $\sim 0.2ev_F/L$ to $\sim 2ev_F/L$, which is one-to-two orders more than predicts the formula $I^{theor}_{typ} \simeq (ev_F/L)(l/L)$.
This disagreement has sofar not been explained, specifically, the effort to explain it by considering the electron-electron interaction (reviewed e.g. in \cite{Saminadayar}) was not successful. Ten years after the work \cite{Chand}, the same laboratory \cite{Jariwala} prepared a new Au samples and observed the typical persistent currents much closer to the formula $I^{theor}_{typ} \simeq (ev_F/L)(l/L)$, but still two-to-three times larger.

On the other hand, recent measurements of the persistent current in a single ring \cite{Bluhm}, performed for thirty Au rings, have shown a good agreement with the formula \eqref{Igre}. In addition, the persistent current in a single Al ring has quite recently been measured by a new highly-sensitive method \cite{Bles}. This work definitely demonstrates agreement of the experimental data
with the formula \eqref{Igre}, modified by a temperature-dependent factor. The experiments \cite{Bluhm,Bles} thus strongly suggest, that the typical persistent current in a single disordered normal-metal ring is not affected by the electron-electron interaction at least for such metals like Au and Al. If this is the case, then the disagreement between the formula \eqref{Igre} and previous measurements of the Au rings \cite{Chand,Jariwala} is not due to the electron-electron interaction and the explanation, if any, may be hidden in the single-particle interaction with disorder.

The formula \eqref{Igre} holds for disorder modeled by the white-noise-like potential with spatially homogenous randomness. In reality, fabrication of the metallic wires/rings from such metals like Au, Ag, Cu, etc., involves techniques like
the electron beam lithography, lift-off, and metal evaporation, which provide wires/rings with disorder due the grain boundaries, impurity atoms and rough edges \cite{Saminadayar}. Thus it seems reasonable to study realistic disorder and to compare the results with the white-noise-based theory \cite{Cheung,riedel}.

In this work, electron transport in the mesoscopic wires and rings is studied with the aim to compare an impurity-free system with grain boundaries with a grain-free system containing the impurity disorder. (We ignore the edge roughness which is studied elsewhere \cite{feilhmosk,feilhauer}.) The mesoscopic wire is called quasi-one-dimensional (Q1D) if its length $L$ is much larger than the width ($W$) and thickness ($H$) \cite{Mello-book}. We mainly study the Q1D wires and rings made of a two-dimensional conductor ($H \rightarrow 0$) of width $W$ and length $L \gg W$, when the dimensionality entering the formula \eqref{Igre} is $d = 2$ \cite{riedel}. At the end we extend our study to the case $d = 3$, i.e., to the
three-dimensional (3D) conductor with $H \sim W$.

\begin{figure}[t]
\centerline{\includegraphics[clip,width=1.0\columnwidth]{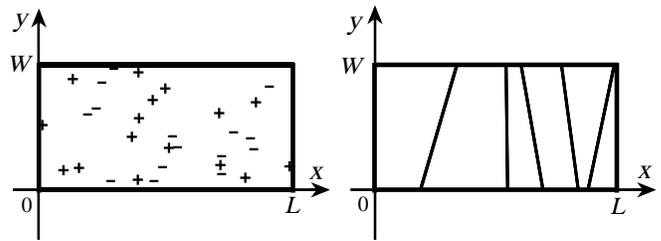}}
\vspace{-0.2cm} \caption{Wire made of the 2D conductor of width $W$ and length $L$. The figure on the left depicts the wire with impurities positioned at random with random signs of the impurity potentials. The figure on the right depicts the impurity-free wire with grain boundaries. The grain boundaries are represented by the straight lines which intersect the wire with a random orientation in randomly chosen positions.}
\label{Fig:1}
\end{figure}

In our $d = 2$ study, a single grain boundary is modeled as a set of the two-dimensional-$\delta$-function-like barriers positioned equidistantly on a straight line and disorder is emulated by a large number of such straight lines, intersecting the conductor with random orientation in random positions (figure \ref{Fig:1}). The impurity
disorder is represented by many two-dimensional
$\delta$-barriers with randomly chosen positions
and signs.
The electron transmission through the wires is calculated by
the scattering-matrix method \cite{tamura,cahay}, and the Landauer conductance is obtained.
To calculate the persistent current in the rings with magnetic flux, we include into the scattering-matrix method the
flux-dependent cyclic boundary conditions and we introduce a trick allowing to study the typical persistent current in rings of almost realistic size.
We mainly focus on the systems with $L \gg l$, when the transport is diffusive.

If the grain boundaries are weakly reflecting, the systems with grain boundaries show for large enough $L$ the same (mean) conductance and the same (typical) persistent current as the systems with impurities. The obtained results also agree with the single-particle theories \cite{Datta-kniha,Cheung,riedel} treating disorder as a white noise.

 If the grain boundaries are strongly reflecting, the rings with the grain boundaries are found to exhibit the typical persistent currents
 which can be (in the diffusive regime) about three-to-four times larger than the white-noise-based result $I^{theor}_{typ} \simeq (ev_F/L)(l/L)$. This finding resembles the experimental findings of reference \cite{Jariwala}.

 Finally, we extend our study to the 3D conductors with the columnar grains \cite{Thompson,Thompson2,Thompson3,Liu,Mazor,Faurie,Harris,Thornton,Yeager,Miller}, which are fundamentally different from the tiny randomly-oriented grains, implicitly assumed in any white-noise-based description of disorder.
We show that the typical persistent current in the diffusive metallic ring with the columnar grains is given by the formula
$I_{typ} \simeq 1.26 \sqrt{N_H} (ev_F/L)(l/L)$, where $N_H \simeq Hk_F/\pi$ is the number of the 2D subbands within the thickness $H$.
For the Au ring with $H = 70$nm  the formula gives the result
 $I_{typ} \simeq 20 (ev_F/L)(l/L)$, which is not far from the experimental results of reference \cite{Chand}.

 In section II we discuss our calculation of the Landauer conductance:
 we review the scattering-matrix method for the wire with impurity disorder and we include the grain boundaries. In section III we describe our calculation of the persistent current. Our results are discussed in section IV, which also contains extension to the case $d=3$. The appendices A, B, and C describe a few technical aspects.

\section{II. Scattering matrix and conductance}

We consider the electron gas confined in the two-dimensional (2D) conductor depicted in the figure \ref{Fig:1}. At zero temperature, the wave function $\varphi(x,y)$ of the electron at the Fermi level ($E_F$) is described by the \schre equation
\begin{equation}
H\varphi(x,y)=E_F\varphi(x,y)
\label{schrodgen}
\end{equation}
with Hamiltonian
\begin{equation}
 H = - \frac{\hbar^2}{2m} \left( \frac{\partial^2}{\partial x^2} + \frac{\partial^2}{\partial y^2} \right) + V \left(y \right)
+ U_D \left( x,y \right) ,
\label{hamiltdisord}
\end{equation}
where $m$ is the electron effective mass, $U_D(x,y)$ is the potential due to disorder, and $V(y)$ is the confining potential due to the edges. The confining potential can be expressed as
\begin{eqnarray}
V(y) = \left\{
                          \begin{array}{ll}
                            0, & 0<y<W \\
                            \infty, & \mbox{elsewhere}
                          \end{array} \right.
.
\label{smoothpot}
\end{eqnarray}
For the impurity disorder we use the simplest model potential
\begin{equation}
U_D(x,y) = \sum_{i} \gamma \delta(x - x_i) \delta(y - y_i),
\label{delta1}
\end{equation}
where we sum over the random impurity positions $[x_i,y_i]$ with a random sign of the impurity strength $\gamma$, as it is shown in the figure \ref{Fig:1}. Disorder due to the grain boundaries can also be modeled by means of \eqref{delta1},
  if the individual $\delta$-barriers in equation \eqref{delta1} are positioned on the straight lines (grain boundaries in the figure \ref{Fig:1})
  equidistantly and with a positive sign of the constant $\gamma$. Details will be given later on. Now it is important that both the impurity disorder and grain-boundary disorder are represented by a sum of the two-dimensional $\delta$-functions. This allows
  us to treat both of them by a very similar scattering-matrix technique. We first review the scattering-matrix technique for the impurity disorder \cite{tamura,cahay,feilhmosk}.

Assume that the disordered wire in figure \ref{Fig:1} is connected to two ballistic semiinfinite contacts of constant width $W$. In the contacts
the electrons obey the \schre equation
\begin{equation}
 \left[ - \frac{\hbar^2}{2m} \left( \frac{\partial^2}{\partial x^2} + \frac{\partial^2}{\partial y^2} \right) + V \left( y \right)
 \right] \varphi(x,y)=E_F\varphi(x,y),
 \label{schrodka}
\end{equation}
where $V(y)$ is the confining potential given by equation \eqref{smoothpot}.
Solving equation \eqref{schrodka} one finds the independent solutions
\begin{equation} \label{subpasovafunkcia}
\varphi_n^\pm (x,y) = \\ e^{\pm ik_nx} \chi_n(y), \quad n = 1,2, \dots \infty,
\end{equation}
with the wave vectors $k_n$ given by equation
\begin{equation}
E_F = \epsilon_n + \frac{\hbar^2 k_n^2}{2m}, \quad \epsilon_n \equiv \frac{\hbar^2 \pi^2}{2mW^2} n^2,
\label{spasy}
\end{equation}
where $\epsilon_n$
is the energy of motion in the $y$-direction and
\begin{equation}
\begin{array}{c}
\chi_n(y) = \left\{
                          \begin{array}{ll}
                            \sqrt{\frac{2}{W}} \sin \left( \frac{\pi n}{W}y \right), & 0<y<W \\
                            0, & \mbox{elsewhere}
                          \end{array} \right.
\end{array}
\label{subbandfunction}
\end{equation}
is the wave function in the direction $y$.
The vectors $k_n$
in \eqref{subpasovafunkcia} are assumed to be positive. The energy $\epsilon_n + \hbar^2 k_n^2/2m$ is called the $n$-th energy channel.
The channels with $\epsilon_n < E_F$ are conducting while the
channels with $\epsilon_n > E_F$ are evanescent.

We define
$A^{\pm}_n(x) \equiv a^{\pm}_{n} e^{\pm i k_n x}$
and
$B^{\pm}_n(x) \equiv b^{\pm}_{n} e^{{\pm} i k_n x}$, where $a^\pm_n$ and $b^\pm_n$ are the amplitudes of the waves moving in the positive and negative directions of the $x$ axis, respectively. The wave function $\varphi(x,y)$ in the contacts
can be expanded in the basis of the eigenstates \eqref{subpasovafunkcia}.
At the boundary $x=0$
\begin{equation}
\varphi(0,y) = \sum\limits^N_{n=1} \left[ A^+_n(0) + A^-_n(0) \right] \chi_n (y),
\label{rozvojdrs1 0}
\end{equation}
while  at the boundary $x=L$
\begin{equation}
\varphi(L,y) = \sum\limits^N_{n=1} \left[ B^+_n(L) + B^-_n(L) \right] \chi_n (y).
\label{rozvojdrs2 L}
\end{equation}
where $N$ is the considered number of channels (ideally $N = \infty$).
We define the vectors $\vc{A}^\pm(0)$ and $\vc{B}^\pm(L)$ with components $A^\pm_{n=1, \dots N}(0)$ and $B^\pm_{n=1, \dots N}(L)$, respectively,
and we simplify the notations $\vc{A}^\pm(0)$ and $\vc{B}^\pm(L)$ as $\vc{A}^\pm$ and $\vc{B}^\pm$.
The amplitudes $\vc{A}^\pm$ and $\vc{B}^\pm$ are related through the matrix equation
  \begin{equation}
\left(
\begin{array}{c}
\vc{A}^- \\
\vc{B}^+ \\
\end{array}
\right)
=
\left[
\begin{array}{cc}
r & t' \\
t & r' \\
\end{array}
\right]
\left(
\begin{array}{c}
\vc{A}^+ \\
\vc{B}^- \\
\end{array}
\right),
\label{S-matrix-rovnica}
\end{equation}
where
\begin{equation}
S
\equiv
\left[
\begin{array}{cc}
r & t' \\
t & r' \\
\end{array}
\right]
\label{S-matica}
\end{equation}
is the scattering matrix. Its dimensions are $2N \times 2N$ and its elements $t$, $r$, $t'$, and $r'$ are the matrices with dimensions $N \times N$. Physically, $t$ and $t'$ are the transmission amplitudes of the waves $\vc{A}^+$ and $\vc{B}^-$, respectively, while $r$ and $r'$ are the corresponding reflection amplitudes. The matrix $t$ is composed of the elements $t_{mn}(k_n)$, where $t_{mn}(k_n)$ is the probability amplitude for transmission from the channel $n$ to channel $m$.

If we know $t_{mn}(k_n)$, the conductance $g$ can be obtained from the Landauer formula \cite{Landauer}.
In units $2e^2/h$ it reads
\begin{equation}
g = \sum\limits^{N_c}_{n=1} T_{n}  = \sum\limits^{N_c}_{n=1} \sum\limits^{N_c}_{m=1} |t_{mn}|^2 \frac{k_m}{k_n},
 \label{Land}
\end{equation}
where $T_n$ is the transmission probability through disorder for the electron impinging disorder in the $n$-th conducting channel and we sum over all ($N_c$) conducting channels. To obtain $t_{mn}$, we need to determine the scattering matrix $S$.

Consider two wires $1$ and $2$, described by the scattering matrices
$S_1$ and $S_2$. The matrices are defined as
\begin{equation}
S_1
\equiv
\left[
\begin{array}{cc}
r_1 & t'_1 \\
t_1 & r'_1 \\
\end{array}
\right], \quad S_2
\equiv
\left[
\begin{array}{cc}
r_2 & t'_2 \\
t_2 & r'_2 \\
\end{array}
\right].
\label{S-matica 1 and 2}
\end{equation}
Let
\begin{equation}
S_{12}
\equiv
\left[
\begin{array}{cc}
r_{12} & t'_{12} \\
t_{12} & r'_{12} \\
\end{array}
\right]
\label{S-matica12}
\end{equation}
is the scattering matrix of the wire obtained by connecting the wires $1$ and $2$ in series.
The matrix $S_{12}$ is related to the matrices $S_1$ and $S_2$ through the matrix equations \cite{Datta-kniha}
\begin{equation}
\begin{array}{l}
t_{12} = t_2[I - r'_1r_2]^{-1}t_1, \\
r_{12} = r_1 + t'_1r_2[I - r'_1r_2]^{-1}t_1, \\
t'_{12} = t'_1[I + r_2[I - r'_1r_2]^{-1}r'_1]t'_2, \\
r'_{12} = r'_2 + t_2[I - r'_1r_2]^{-1}r'_1t'_2,
\end{array}
\label{skladka}
\end{equation}
where $I$ is the unit matrix.
The equations \eqref{skladka} are usually written in the symbolic form
\begin{equation}
S_{12} = S_1\otimes S_2 .
\label{compositionlaw2}
\end{equation}

\begin{figure}
\centering
\scalebox{0.3}{
\includegraphics*[0mm,0mm][280mm,150mm]{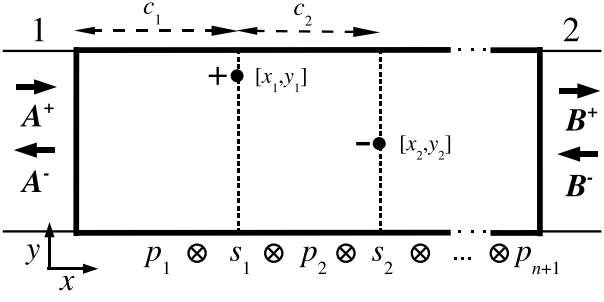}
}
\caption{Wire with the randomly positioned point-like impurities, where an individual impurity is represented by the two-dimensional-$\delta$-function-like potential with random sign. The $n$ impurities described by the scattering matrices $s_i$ divide the wire into the $n+1$ free regions described by the matrices $p_i$. Also shown are the wave amplitudes $\vc{A}^\pm$ and $\vc{B}^\pm$.}
\label{prim}
\end{figure}

Consider the wire with impurity potential \eqref{delta1}. Between any two neighboring impurities there is a region with zero impurity potential, say the region $x_{i-1}<x<x_i$,
where the electron moves along the $x$ axis like a free particle. The wire with $n$ impurities contains $n+1$ regions with free electron motion, separated by $n$ point-like regions where the scattering takes place. As illustrated in figure \ref{prim},
the scattering matrix $S$ of such wire can be obtained by applying the combination law
\begin{equation}
S = p_1\otimes s_1 \otimes p_2 \otimes s_2 \otimes\dots s_n\otimes p_{n+1},
\label{bezprek}
\end{equation}
where $p_i$ is the scattering matrix of free motion in the region $x_{i-1}<x<x_i$ and $s_i$ is the scattering matrix of the $i$-th impurity. The symbols $\otimes$ mean that the composition law \eqref{compositionlaw2} is applied in \eqref{bezprek} step by step: one first combines
the matrices $p_1$ and $s_1$, the resulting matrix is combined with $p_2$, etc.

The scattering matrix $p_i$ can be expressed as
\begin{equation}
p_i
=
\left[
\begin{array}{cc}
0 & \Phi \\
\Phi & 0 \\
\end{array}
\right],
\label{matrixoffree motion}
\end{equation}
where $0$ is the $N \times N$ matrix with zero matrix elements and $\Phi$ is the $N \times N$ matrix with matrix elements
\begin{equation}
\Phi_{mn} = e^{ik_n c_i} \delta_{mn}, \quad c_i = x_i - x_{i-1},
\label{matrixelementoffree motion}
\end{equation}
Finally, the scattering matrix
\begin{equation}
s_i
\equiv
\left[
\begin{array}{cc}
r & t' \\
t & r' \\
\end{array}
\right]
\label{s_i-matica}
\end{equation}
is composed of the matrices
\begin{eqnarray}
t = t' = [K + i\Gamma ]^{-1} K, \label{tt}\\
r = r' = - [K + i\Gamma]^{-1} i \Gamma,
\label{rr}
\end{eqnarray}
where $K$ and $\Gamma$ are the $N \times N$ matrices with matrix elements
\begin{equation}
K_{m n} = k_n \delta_{m n} \ \ , \ \ \Gamma_{m n} = \frac{m\gamma }{\hbar^2} \chi^{*}_{m}(y_i)\chi_n(y_i).
\label{maticky}
\end{equation}

\begin{figure}
\centering
\includegraphics[width=8cm]{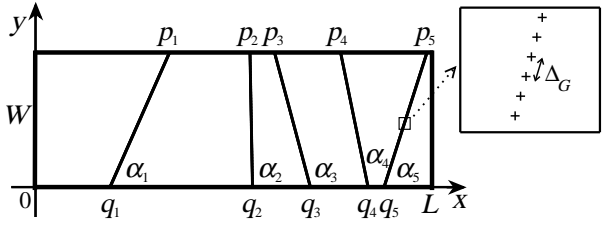}
\caption{Wire with grain boundaries represented by straight lines. The $i$th grain boundary starts at point $q_i$ and ends at point $p_i$. The positions $q_i$ and $p_i$ are chosen as random. The avoid appearance of the mutually intersecting boundaries, the random numbers are ordered increasingly,
i.e., $q_1 < q_2 \dots < q_5$ and $p_1 < p_2 \dots < p_5$. The angle between the $i$-th boundary and wire edge is $\alpha_i$, another important parameter is the mean lateral size of the grain, $d_G$. Inset shows in detail a single boundary. The boundary is represented by a set of the equidistantly-positioned repulsive point-like impurities (plus signs), where a single impurity is modeled as a two-dimensional $\delta$-function-like energy barrier (see the text). If we choose the nearest-neighbor distance $\Delta_{G} \ll \lambda_F$, the grain boundary effectively behaves as a structure-less one-dimensional energy barrier.
}
\label{zrna}
\end{figure}

The scattering matrix method for the grain boundaries is the same like for the impurities, because
a single grain boundary is formally modeled by a set of the point-like impurities (see figure \ref{zrna}). We start with the grain boundaries oriented perpendicularly to the wire. Disorder due to the perpendicular boundaries is modeled by the potential
\begin{equation}
U_D(x,y) = \sum_i \gamma_G \delta(x-x_i),
\label{Uzrnocel}
\end{equation}
where $\gamma_G$ is the strength of the perpendicular boundary and $x_i$ is its random position along the wire.
Obviously, the $S$-matrix of the wire with perpendicular boundaries is given by the combination law \eqref{bezprek}, where  $s_i$
 are the scattering matrices of the individual boundaries and the matrices $p_i$ describe the free electron motion between two neighboring boundaries. The potential of the perpendicular boundary
at $x=0$ reads
\begin{equation}
U_D(x,y) = \gamma_G \delta(x).
\label{Uzrno}
\end{equation}
Formally, it is a one-dimensional version of the impurity potential $\gamma \delta(x) \delta(y - y_i)$, for which the matrix $s_i$ is known: it is given by equations
 \eqref{tt}, \eqref{rr}, and \eqref{maticky}. Therefore, the scattering matrix $s_i$ for the potential \eqref{Uzrno} is given by the same equations, except that the elements of the matrix $\Gamma$ now read
\begin{equation}
\Gamma_{m n} = \frac{m\gamma_G}{\hbar^2} \delta_{m n}.
\label{Gmaticka}
\end{equation}
The elements of the matrices $t$, $t'$, $r$, and $r'$ can be written as
\begin{equation}
t_{m n} = t'_{m n} = \frac{k_n}{k_n + i\bar{\gamma}_G} \delta_{m n},
\label{tgkol}
\end{equation}
\begin{equation}
r_{m n} = r'_{m n} = -\frac{i\bar{\gamma}_G}{k_n + i\bar{\gamma}_G} \delta_{m n},
\label{rgkol}
\end{equation}
where $\bar{\gamma}_G  = m \gamma_G / \hbar^2$. Since the matrices $t$, $t'$, $r$, and $r'$ are diagonal, the electron impinging the perpendicular boundary in the channel $n$ is reflected back to the same channel. The reflection probability for the channel $n = 1$  is
\begin{equation}
R_G \equiv |r_{11}|^2 = \frac{\bar{\gamma}^2_G}{k^2_F + \bar{\gamma}^2_G},
\label{Rgv}
\end{equation}
where we use the approximation $k_1 \simeq k_F$, with $k_F$ being the 2D Fermi wave vector. In other words, the 2D electron
impinging the grain boundary perpendicularly is reflected from $k_F$ to $-k_F$ with the reflection probability $R_G$ coinciding with $|r_{11}|^2$.
The equation \eqref{Rgv} allows us to describe the grain boundary by the parameter $R_G$ which is measurable.

In real metallic wires the perpendicular grain boundaries usually do not exist \cite{Bietsch}. Indeed,
 the matrix elements \eqref{tgkol} and \eqref{rgkol} are diagonal. This means that there is no inter-channel scattering, i.e., transport through such wire takes place in the mutually independent channels. However, a single disordered 1D channel is always in the localization regime \cite{markos,MoskoPRL} while the metallic Q1D wires usually exhibit diffusive regime \cite{Mohanty}. So we consider the grain boundaries with random orientation.

As shown in the figure \ref{zrna}, a single grain boundary is modeled by a set of the equidistantly-positioned repulsive point-like impurities with the nearest-neighbor distance $\Delta_{G} \ll \lambda_F$.
In this model, the potential of the grain boundaries is given as
\begin{equation}
U_D(x,y) = \sum_{i} B_i(x,y),
\label{Gtodelta}
\end{equation}
where
\begin{equation}
B_i(x,y) = \sum_{y_j < W} \gamma \delta(x - x_{ij}) \delta(y - y_{ij}),
\label{Gtodelta1}
\end{equation}
is the potential of the $i$th boundary, the same positive $\gamma$ is used for all impurities, and $[x_{ij},y_{ij}]$ is the position of the $j$th impurity at the $i$th boundary. Following the figure \ref{zrna} we find
\begin{equation}
x_{ij} = q_i + j\Delta_G \cos(\alpha_i), \quad y_{ij} = j \Delta_G \sin(\alpha_i).
\label{Gtodeltaxy}
\end{equation}
The grain boundaries described by the potential \eqref{Gtodelta}-\eqref{Gtodeltaxy} are formally a special
case of the impurity disorder and therefore can be treated by the same scattering-matrix algorithm.

 In our model, the reflectivity of a single randomly-oriented grain boundary depends on the parameters $\Delta_G$ and $\gamma$.
 We can find the relation between these parameters and parameter $R_G$, defined by equation \eqref{Rgv}.
  Assume that the grain boundary described by potential \eqref{Gtodelta1} intersects the wire perpendicularly at $x=0$. This simplifies \eqref{Gtodelta1} into the form
\begin{equation}
B(x,y) = \sum_{j} \gamma \delta(x) \delta(y - j\Delta_G).
\label{Gtodeltakolm}
\end{equation}
Here $\gamma \delta(x) \delta(y - j\Delta_G)$ is the same single-impurity potential, for which we have already expressed the scattering matrix (equations \ref{rr}, \ref{tt}, and \ref{maticky}). Therefore, the matrix $\Gamma$ of the potential \eqref{Gtodeltakolm}
is simply a sum of the $\Gamma$ matrices of all individual potentials $\delta(x) \delta(y - j\Delta_G)$, i.e.,
\begin{equation}
\Gamma_{m n} = \frac{m}{\hbar^2} \frac{\gamma}{\Delta_G} \sum_{y_i < W} \chi^{*}_{m}(y_{i-1}+\Delta_G)\chi_n(y_{i-1}+\Delta_G) \Delta_G.
\label{Gamtod}
\end{equation}
For small $\Delta_G$ the sum in the equation \eqref{Gamtod} can be replaced by integral $\int^W_0 \chi^*_{m}(y) \chi_n(y) dy$
and we obtain
\begin{equation}
\Gamma_{m n} = \frac{m}{\hbar^2} \frac{\gamma}{\Delta_G} \delta_{m n}.
\label{Gamtodf}
\end{equation}
Comparing this expression with \eqref{Gmaticka} we obtain the relation $\gamma_G = \gamma/\Delta_G$. The perpendicular reflectivity \eqref{Rgv} becomes
\begin{equation}
R_G = \frac{\bar{\gamma}^2}{k^2_F \Delta^2_G + \bar{\gamma}^2},
\label{Rgvdelta}
\end{equation}
where $\bar{\gamma} = m \gamma / \hbar^2$. The randomly-oriented grain boundaries can thus be characterized by a single parameter $R_G$, related
to the model parameters $\bar{\gamma}$ and $\Delta_G$ through the equation \eqref{Rgvdelta}. If we use $\Delta_{G} \ll \lambda_F$,
the resulting wire conductance (for a fixed value of $R_G$) is independent on the choice of $\bar{\gamma}$ and $\Delta_G$.

\section{III. Calculation of persistent current}

\begin{figure}
\centering
\includegraphics[width=6cm]{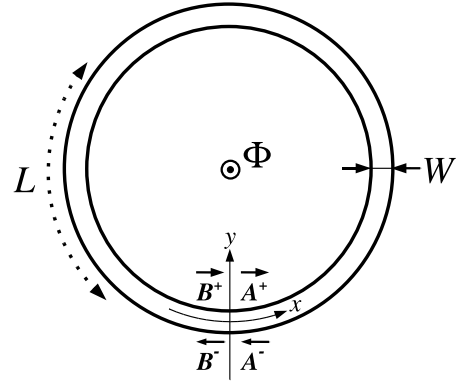}
\caption{Circular ring of width $W$ and length $L \gg W$, pierced by magnetic flux $\phi$. Also shown are the wave amplitudes $\vc{A}^\pm$ and $\vc{B}^\pm$ entering the scattering-matrix equation }
\label{prstenec}
\end{figure}

We consider a circular ring of width $W$ and length $L \gg W$, shown in the figure \ref{prstenec}. The opening of the ring is pierced by magnetic flux $\phi$ due to the magnetic field directed along the axis $z$. The ring is in fact the Q1D wire from the previous text, but circularly shaped and with the wire ends connected. Therefore, the electron wave function $ \varphi(x,y)$ and electron energy $E$ in the ring can still
be described by the
\schre equation \eqref{schrodgen} with Hamiltonian \eqref{hamiltdisord}, but we also need
to ensure the continuity of the wave function and its first derivative at the connection. This implies the boundary conditions
\begin{equation}
\begin{array}{c}
\varphi_n(0,y) = \exp \left(- i 2 \pi \frac{\phi}{\phi_0} \right) \varphi_n(L,y), \\
 \frac{\partial \varphi_n}{\partial x} (0,y) = \exp \left(- i 2 \pi \frac{\phi}{\phi_0} \right) \frac{\partial \varphi_n}{\partial x} (L,y).
\end{array}
\label{hrpo}
\end{equation}
where $\phi_0 = h/e$ is the flux quantum and the exponential factor is the Peierls phase factor due to the flux $\phi$.
Due to the boundary conditions \eqref{hrpo} the energy $E$ is discrete and depends on $\phi$. Now we show how to find the spectrum $E_n(\phi)$ \cite{feilhauer}.

\begin{figure}
\centering
\includegraphics[width=8cm]{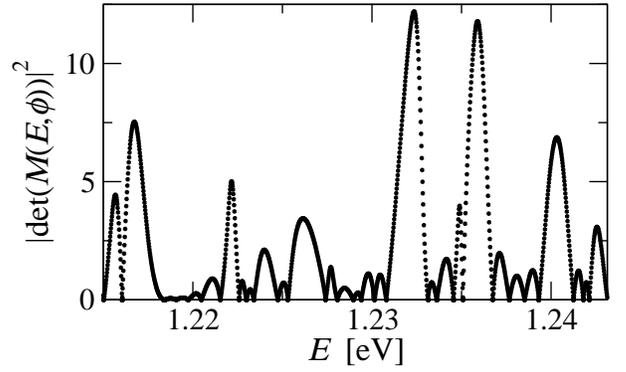}
\caption{Determinant $|\mbox{det}(M(E,\phi))|^2$ versus $E$. The presented data points are the values of $|\mbox{det}(M(E,\phi))|^2$, calculated for equidistant energies with a small energy step $\Delta E$. The minima of the sharp valleys are the zero values of $|\mbox{det}(M(E,\phi))|^2$): each zero occurs at an eigen-energy $E_n(\phi)$. All data were obtained for the specific ring parameters discussed in figure \ref{Fig-5}, but
the presented dependence is (qualitatively) typical for any disordered ring.}
\label{detok}
\end{figure}

Since we describe the ring by equations \eqref{schrodgen} and \eqref{hamiltdisord}, we can directly
apply the scattering-matrix method developed in the preceding text. Indeed,
 the wave function $ \varphi(x,y)$ can be expressed in the ring positions $x=0$ and $x=L$ by means of the expansions \eqref{rozvojdrs1 0}
and \eqref{rozvojdrs2 L}, where the amplitudes $\vc{A}^\pm$ and $\vc{B}^\pm$ are related through the scattering-matrix equation \eqref{S-matrix-rovnica}. If we set the expansions \eqref{rozvojdrs1 0} and \eqref{rozvojdrs2 L} into the boundary conditions \eqref{hrpo},
we can rewrite  \eqref{hrpo} into the matrix form
\begin{equation}
\left(
\begin{array}{c}
\vc{A}^- \\
\vc{B}^+ \\
\end{array}
\right)
=
\left[
\begin{array}{cc}
0 & Q^{-1}(\phi) \\
Q(\phi) & 0 \\
\end{array}
\right]
\left(
\begin{array}{c}
\vc{A}^+ \\
\vc{B}^- \\
\end{array}
\right),
\label{okrajka}
\end{equation}
where $Q_{\alpha \beta}(\phi) = \exp(i 2\pi \phi/\phi_{0}) \delta_{\alpha \beta}$.
Combining the matrix equations \eqref{S-matrix-rovnica} and \eqref{okrajka} we obtain the equations
\begin{equation}
\left[
\begin{array}{cc}
0 & Q^{-1}(\phi) \\
Q(\phi) & 0 \\
\end{array}
\right]
\left(
\begin{array}{c}
\vc{A}^+ \\
\vc{B}^- \\
\end{array}
\right)
=
\left[
\begin{array}{cc}
r & t' \\
t & r' \\
\end{array}
\right]
\left(
\begin{array}{c}
\vc{A}^+ \\
\vc{B}^- \\
\end{array}
\right),
\end{equation}
which can be rearranged into the form
\begin{equation}
\left[
\begin{array}{cc}
r & t' - Q^{-1}(\phi) \\
t - Q(\phi) & r' \\
\end{array}
\right]
\left(
\begin{array}{c}
\vc{A}^+ \\
\vc{B}^- \\
\end{array}
\right)
=0.
\label{finalka}
\end{equation}
We label the matrix on the left side of \eqref{finalka} as $M(E,\phi)$. To fulfill the equation \eqref{finalka}, the determinant of the matrix $M(E,\phi)$ has to be zero, i.e.,
\begin{equation}
\mbox{det}(M(E,\phi)) = \mbox{det}
\left[
\begin{array}{cc}
r(E) & t'(E) - Q^{-1}(\phi) \\
t(E) - Q(\phi) & r'(E) \\
\end{array}
\right]
=0.
\label{detac}
\end{equation}
The submatrices $t$, $r$, $t'$ and $r'$ are functions of the electron energy $E$. Therefore, the matrix $M$ is a function of the magnetic flux $\phi$ and energy $E$. The determinant of $M(E,\phi)$ is a complex number. Therefore, the real as well as imaginary parts of $\mbox{det}(M(E,\phi))$ have to
be zero to fulfill the equation \eqref{detac}. The equation \eqref{detac} is thus equivalent to the equation
\begin{equation}
|\mbox{det}(M(E,\phi))|^2 = 0,
\label{absvaldet}
\end{equation}
which we solve numerically. For a given value of magnetic flux,
the determinant $|\mbox{det}(M(E,\phi))|^2$ is calculated numerically as a function of the energy $E$ which is varied with a small energy step $\Delta E$ from zero up to the Fermi energy. In the figure \ref{detok}, a typical numerical result for $|\mbox{det}(M(E,\phi))|^2$ is shown for a small energy window. The eigen-energies $E_n(\phi)$ are the zero points of $|\mbox{det}(M(E,\phi))|^2$. In the figure \ref{detok}
the oscillating function $|\mbox{det}(M(E,\phi))|^2$ show a series of very sharp valleys with a zero minimum value, i.e., the positions of these minima are the eigen-energies of interest.
We repeat this procedure for magnetic flux $\phi + \Delta \phi$, where $\Delta \phi = 10^{-4} \phi_0$, and we obtain the eigen-energies $E_n(\phi + \Delta \phi)$.

At zero temperature the persistent current is given as \cite{buttiker83,byers}
\equr{pc}{I = \sum_{ \forall E_n \leq E_F} I_n = -\sum_{\forall E_n \leq E_F} \frac{\dif E_n}{\dif \phi},}
where one sums the single-electron currents for all energy levels below the Fermi level.
Thus, to obtain the persistent current in a single sample, one needs to determine all single-electron energies
$E_n(\phi + \Delta\phi)$ and $E_n(\phi)$ below the Fermi level, to evaluate all single electrons currents
\begin{equation}
I_n(\phi) = - \frac{\partial E_n}{\partial \phi} (\phi) \simeq - \frac{E_n(\phi + \Delta\phi) - E_n(\phi)}{\Delta\phi}.
\label{numder}
\end{equation}
and to sum them as shows the equation \eqref{pc}. This procedure is computationally cost and allows to study only small rings,
but later on we introduce a trick allowing to study large rings.

Finally, we note that describing the ring by means of \eqref{schrodgen}, \eqref{hamiltdisord}, and \eqref{hrpo} we ignore the magnetic field in the ring.
Further, the approach ignores the effect of the ring curvature, because we assume that the $x$ axis circulates along the ring circumference (see figure \ref{prstenec}). Both approximations hold for $L \gg W$.

\section{IV. Results}

\subsection{A. Conductance of wires with grain boundaries and impurities}

In this subsection we present our scattering-matrix results for the wire conductance.
The wires with disorder due to the randomly-oriented grain boundaries are compared with the wires with impurity disorder. Also included are the wires with disorder due to the perpendicular grain boundaries.

We use the material parameters $m=9.109\times10^{-31}$kg and $E_F = 5.6$eV ($\lambda_F = 0.52$nm), typical of the Au wires. We first study the Au wires of width $W = 9$nm, with
the number of the conducting channels being $N_c = 34$. This number well emulates the limit $N_c \gg 1$, but later we also use larger $N_c$.

The parameters of the grain-boundary disorder are chosen as follows. The perpendicular reflectivity $R_G$ (equation \ref{Rgvdelta}) and the mean lateral size of the grain, $d_G$, are kept the same for the randomly-oriented as well as perpendicular boundaries in order to isolate the effect of random orientation. We recall that a single grain boundary is modeled as a line with equidistant impurities of strength $\bar{\gamma}$ and nearest-neighbor distance $\Delta_G$. Since we keep $\Delta_G \ll \lambda_F$,  the choice of $\bar{\gamma}$ and $\Delta_G$ has no effect on the resulting conductance for a given $R_G$. However, once the parameters  $d_G$, $\Delta_G$ and $\bar{\gamma}$ are chosen, we keep the same $\bar{\gamma}$ and the same total number of impurities also in the wire with impurity disorder. Both types of disorder are thus represented by the same numerical model. Therefore, any difference between their transport properties reflects exclusively the difference between the scattering by repulsive lines and scattering by a random array of point-like scatterers.

 In the ensemble of macroscopically identical wires disorder fluctuates from wire to wire and so does the conductance. Hence we evaluate \eqref{Land} for the ensemble of (typically)  $10^3$ wires and we obtain the mean conductance $\langle g \rangle$, mean resistance $\langle \rho \rangle$ where $\rho = 1/g$, and variance $\langle g^2 \rangle - \langle g \rangle^2$.

\begin{figure}[t]
\centerline{\includegraphics[clip,width=1.0\columnwidth]{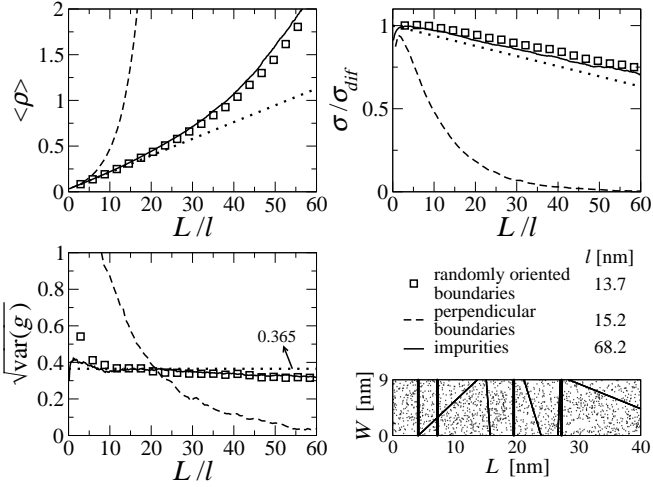}}
\vspace{-0.15cm} \caption{Mean resistance $\langle \rho \rangle$, conductivity $\sigma$, and conductance fluctuations $\sqrt{\mbox{var}(g)}$ versus $L/l$. The results for the wires with impurity disorder, randomly-oriented grain boundaries, and perpendicular grain boundaries are
shown by full lines, squares, and dashed lines, respectively.
The dotted line in the top left panel is the linear fit $\langle \rho \rangle = 1/N_c + \rho_{dif}L/W$, where
$\rho_{dif} = 2/(k_F l)$ is the diffusive resistivity and the mean free path $l$ is fitted.
The conductivity $\sigma$ is extracted from the mean conductance $\langle g \rangle$ by means of \eqref{sigmag} and normalized by $\sigma_{dif}=1/\rho_{dif}$. The dotted line in the top right panel shows the theoretical formula \eqref{sigmaweaksc}. The parameters of the grain-boundary disorder are $R_G = 0.2$ and $d_G = 10$nm, the impurity strength is $\bar{\gamma} = 0.303$, and the impurity density is $n_I = 2.18nm^{-2}$. All three types of disorder are shown schematically in the right bottom panel.
}
\label{trans}
\end{figure}

The figure \ref{trans} shows our results for the mean resistance, conductance, and conductance fluctuations in dependence on the wire length. If we compare the wires with randomly-oriented grain boundaries (data shown by squares) with the wires with impurity disorder (data shown by the full lines), we see that the results for both types of wires are in good mutual agreement and also in accord with what one expects for the white-noise-like disorder. The following features are worth to stress.

 First we look at the mean resistance. Both the impurity disorder and randomly-oriented grain boundaries first show the linear diffusive dependence \cite{Datta-kniha}
\begin{equation}
\langle \rho \rangle = \frac{1}{N_c} + \rho_{dif}\frac{L}{W},  \quad \rho_{dif} = \frac{k_F}{\pi n_e l},
\label{rhoweaklin}
\end{equation}
where $\rho_{dif}$ is the diffusive resistivity and $n_e = k_F^2/2 \pi$ is the 2D electron density. Notice that in the former case $l = 68$nm while in the latter case $l = 13.7$nm only. In other words, the point-like scatterers constituting the repulsive lines scatter the electrons much more effectively like the point-like scatterers of the equivalent strength in a random lattice.

For $L \gg l$ the full line and squares start to deviate from the linear rise \eqref{rhoweaklin}. The deviation is due to the weak localization and eventually due to the strong one, manifested by exponential rise of $\langle \rho \rangle$ with $L$.
On the other hand, for the wire with
perpendicular grain boundaries (dashed line) we see the exponential rise of $\langle \rho \rangle$ already for $L/l \sim 1$, which means that the Q1D wire is in the localization regime. This is because each channel behaves like an independent 1D disordered channel.

The figure \ref{trans} also shows the wire conductivity
\begin{equation}
\sigma = \frac{1}{1/\langle g \rangle - 1/N_c} \frac{L}{W},
\label{sigmag}
\end{equation}
normalized by the diffusive conductivity $\sigma_{dif}=1/\rho_{dif}$. In absence of localization
$\sigma/\sigma_{dif} = 1$ independently on $L$. In fact, we see that $\sigma/\sigma_{dif}$
decreases with $L$ linearly both for the impurity disorder and randomly-oriented grain boundaries. This linear
decrease is in accord with the weak-localization-mediated behavior predicted for the white-noise-like disorder \cite{Mello-book,MelloStone}.
Indeed, the theory \cite{Mello-book,MelloStone} predicts
\begin{equation}
\langle g \rangle = \sigma_{dif} \frac{W}{L} - \frac{1}{3}, \quad l \ll L \ll \xi,
\label{Gweak unnorm}
\end{equation}
where $\sigma_{dif} W/L$ is the classical diffusion term and the term $1/3$ is the weak localization correction typical of the Q1D wire.
If we write \eqref{Gweak unnorm} in terms of the conductivity, we obtain
\begin{equation}
\frac{\sigma}{\sigma_{dif}} = 1 - \frac{1}{3}\frac{k_F}{\pi n_e W}  \frac{L}{l}.
\label{sigmaweaksc}
\end{equation}
In the figure \ref{trans} this equation is compared with the numerical data for $\sigma/\sigma_{dif}$. Indeed, the agreement is very good both for the impurities and randomly-oriented grain boundaries. On the contrary, for the
perpendicular grain boundaries we see that $\sigma/\sigma_{dif}$ decreases with $L$ exponentially. In such Q1D wire there is no weak localization, only the strong one.

\begin{figure}[t]
\centerline{\includegraphics[clip,width=1.0\columnwidth]{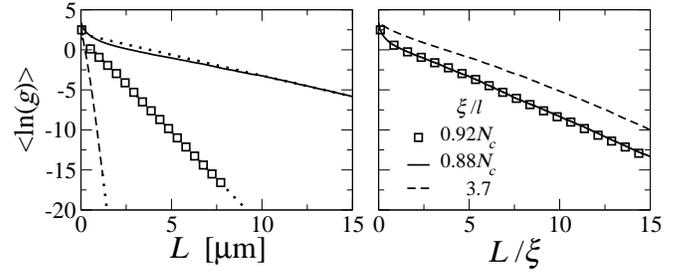}}
\vspace{-0.15cm} \caption{Typical conductance $\langle \ln g \rangle$ in dependence on $L$ and $L/\xi$. The results for the wires with impurity disorder, randomly-oriented grain boundaries, and perpendicular grain boundaries are
shown by full lines, squares, and dashed lines, respectively. The dotted lines show the fit $\langle \ln g \rangle = - L/\xi$ which gives the localization lengths $\xi$ shown in the figure. All parameters are the same as in the figure \ref{trans}.
}
\label{trans1}
\end{figure}

Finally, we look at the conductance fluctuations $\sqrt{\mbox{var}(g)} \equiv \sqrt{\langle g^2 \rangle - \langle g \rangle^2}$.
For the Q1D wire with the white-noise-like disorder the theory predicts the universal value \cite{LeeStone,fukuyama}
\begin{equation}
\sqrt{\mbox{var}(g)} = 0.365.
\label{varweak}
\end{equation}
The figure \ref{trans} shows that the impurity disorder and randomly-oriented grain boundaries
exhibit $\sqrt{\mbox{var}(g)}$ in accord with prediction \eqref{varweak}. For the perpendicular boundaries
 we see a quite different $\sqrt{\mbox{var}(g)}$ as the diffusive regime is absent.

The figure \ref{trans1} shows
the typical conductance $\langle \ln{g} \rangle$ versus the wire length. For all three types of disorder, our numerical data approach at large $L$ the dependence
$\langle \ln{g} \rangle = - L/\xi$ \cite{LeeStoneAnderson,fukuyama}. This is a sign of the localization. Fitting of the numerical data provides the values of $\xi$ shown in the figure.
 We find the result $\xi/l \simeq 0.9 N_c$ for the impurity disorder as well as for the randomly-oriented grain boundaries. The result $\xi/l \simeq 0.9 N_c$ reasonably agrees with the result $\xi/l = N_c$ predicted for the white-noise-like disorder \cite{Thouless} and with the numerical studies for impurity disorder \cite{tamura}. For the perpendicular grain boundaries we find the value $\xi/l \simeq 3.7$, which differs from the 1D result $\xi/l_{1D} =1$ \cite{Datta-kniha}. The difference is due to the fact that our $l$ is the mean over many channels.

  \begin{figure}[t]
\centerline{\includegraphics[clip,width=1.0\columnwidth]{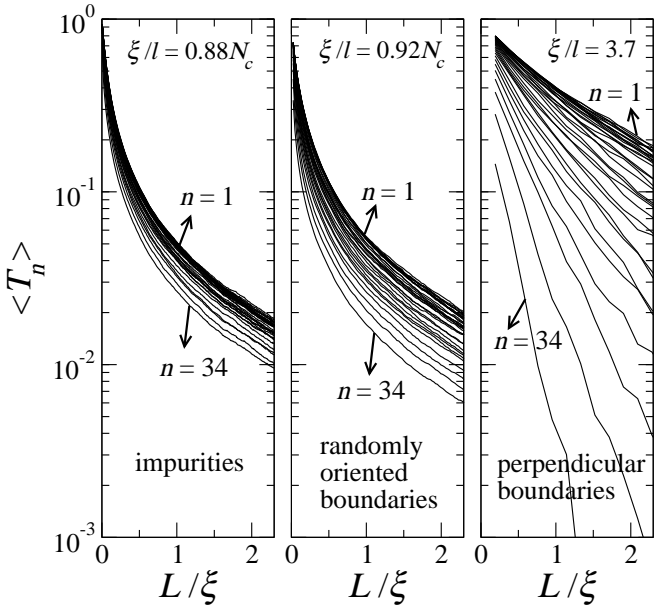}}
\vspace{-0.15cm} \caption{Transmission probability $\langle T_n \rangle$ versus $L/\xi$ for the channel indices $n =1, 2, \dots N_c$, where $N_c = 34$. For $n$ ordered increasingly, the resulting curves are ordered
 decreasingly: the top curve shows $\langle T_{n=1} \rangle$, the bottom one shows $\langle T_{n=N_c} \rangle$. The data are presented separately for the impurity disorder, randomly-oriented grain boundaries, and perpendicular grain boundaries. All presented data originate from the same calculation as the data in figure \ref{trans}.}
\label{Tickas}
\end{figure}

The figure \ref{Tickas} shows the numerical data for $\langle T_n \rangle$. The theory based on the white-noise disorder predicts, that the conducting channels are equivalent \cite{Beenakker,markos}
in the sense that
$\langle T_1 \rangle = \langle T_2 \rangle \dots = \langle T_{N_c} \rangle$. In the figure \ref{Tickas}, this equivalency is reasonably confirmed for the wire with impurity disorder and for the wire with randomly-oriented grain boundaries. Nevertheless, in the latter case the equivalency is not so good as in the former one. This can be understood if we look at the sketch of the grain boundaries in figure \ref{zrna}. It is obvious that the boundaries with the angles $\alpha \rightarrow 0$ or $\alpha \rightarrow \pi$ are very unlikely because
the mutually intersecting boundaries are prohibited. Consequently, the probability distribution of $\alpha$ in the interval $(0, \pi)$
is not homogenous: it has a broad maximum around $\pi/2$. Our scattering-matrix approach works also for the intersecting boundaries, but such study is beyond the scope of this paper: In such case the angle distribution tends to be homogenous in the whole interval $(0, \pi)$, which improves the channel equivalency. Disorder with non-intersecting grain boundaries, studied here, is typical for the so-called bamboo-like wires \cite{Arzt,Austin,WebbWashburn,Neuner}, with $d_G > W$. Finally, for the perpendicular boundaries the channel equivalency is absent due to the localization.

\begin{figure}[t]
\centerline{\includegraphics[clip,width=1.0\columnwidth]{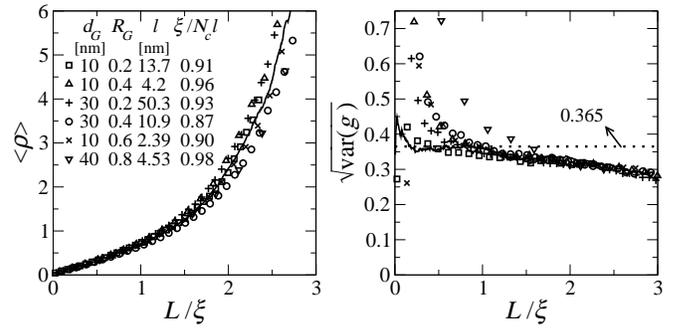}}
\vspace{-0.15cm} \caption{Mean resistance $\langle \rho \rangle$ and conductance fluctuations $\sqrt{\mbox{var}(g)}$ versus $L/\xi$.
The results for the wires with randomly-oriented grain boundaries (shown by symbols) are compared for various values of the reflectivity $R_G$ and grain size $d_G$. The results for the wire with impurity disorder (the same as in figure \ref{trans}) are presented in a full line.}
\label{scaling}
\end{figure}

In the figure \ref{scaling} the wires with the randomly-oriented grain boundaries are studied for various values of the grain-boundary reflectivity $R_G$ and grain size $d_G$. The experimentally measured values of $R_G$ in the polycrystalline wires
range from $0.1$ up to $0.8$ in dependence on the fabrication conditions \cite{durkan,Bietsch,Reiss,Schneider,Sambles,Vries,feldman}.
For all $R_G$ and $d_G$ considered in the figure \ref{scaling}, the resulting mean resistance and conductance fluctuations are similar to the results for impurity disorder (full lines), albeit a noticeable quantitative differences emerge with increasing $R_G$. The data for the conductance fluctuations suggest that
the conductance fluctuations in realistic samples might be correlated with the measured values of $R_G$.
However, \emph{no matter what is the value of $R_G$, the mean resistance of the Q1D wire
with the randomly-oriented grain boundaries rises with $L$ linearly up to $L \simeq \xi$, where $\xi \simeq N_c l$ even for $R_g$ as large as $0.8$.}
To see the standard diffusive regime for the strongly reflecting boundaries is rather surprising.

\subsection{B. Persistent current in rings with grain boundaries and impurities}

In this subsection the persistent currents are studied numerically in the rings with randomly-oriented grain boundaries and rings with impurity disorder. Our numerical results are compared with the theoretical result (equation \ref{Igre}) valid for the diffusive rings with white-noise-like disorder.

In the ensemble of the macroscopically-identical disordered rings the persistent current \eqref{pc} strongly fluctuates from sample to sample.
To asses a typical size of the current in a single sample, one can calculate the typical persistent current
\equr{}{I_{typ} = \sqrt{\av{I^2}},}
where $<\dots >$ means the ensemble averaging. In fact, the persistent current $I$ fluctuates also in a single ring when the number of the electrons (the Fermi energy) is varied. It has been found in \cite{mont} that averaging over the electron number, performed for a single configuration of disorder, leads to the same results as the averaging over different configurations and number of particles at the same time. In this work we average over the electron number (over the Fermi energy) in a single disordered sample. This helps to reduce the computational time, but for our purposes still not sufficiently. Fortunately, we will see soon that the computational time can be further decreased remarkably, when the typical persistent current is studied for magnetic flux $\Phi = \pm 0.25 h/e$. In what follows we use $\Phi = - 0.25 h/e$.

\begin{figure}[t]
\centerline{\includegraphics[clip,width=1.0\columnwidth]{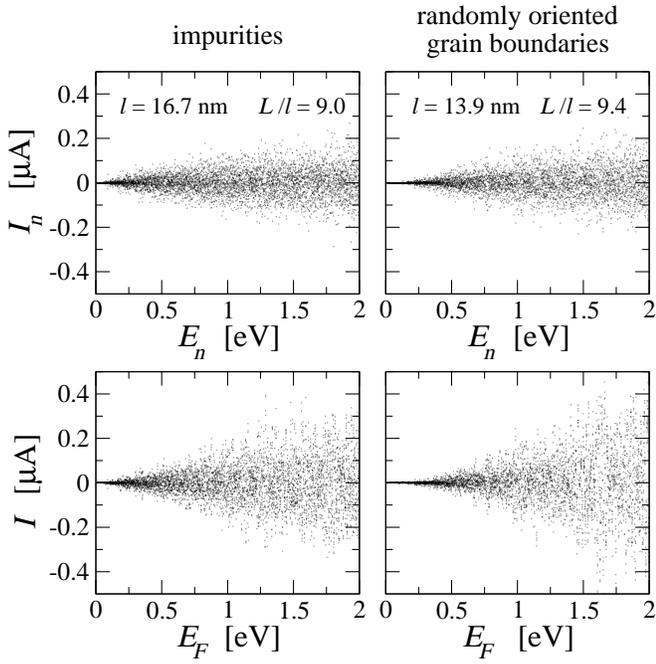}}
\vspace{-0.15cm} \caption{The top panels show the single-electron current $I_n$ versus the eigen-energy $E_n$, calculated by using the relation \eqref{numder} and algorithm described in the figure \ref{detok}. The bottom panels show the persistent current $I(E_F) = \sum_{ \forall E_n \leq E_F} I_n$, where we sum the single-electron currents from the top panels. The dependence $I(E_F)$ is centered around zero mean current, as expected for flux $\Phi = \pm 0.25 h/e$. The parameters of the ring with randomly-oriented grain boundaries are $W=9nm$, $R_G(2eV) = 0.2$, $d_G = 10nm$, $l(2eV) = 13.9nm$, and $L/l = 9.4$. For the ring with impurity disorder $W=9nm$, $n_I=2.5nm^{-2}$, $\bar{\gamma} = 0.4$, $l(2eV) = 16.7nm$, and $L/l = 9.0$.} \label{Fig-5}
\end{figure}

The figure \ref{Fig-5} shows the single-electron current $I_n$ versus $E_n$ and persistent current
$I = \sum_{ \forall E_n \leq E_F} I_n$ versus $E_F$, calculated for the ring with impurity disorder and ring with randomly-oriented
boundaries. In both cases the ring parameters (see the figure caption) are chosen to give roughly the same mean-free path $l$ and ratio $L/l$.
In spite of their chaotic nature, the data for $I(E_F)$ are centered symmetrically around zero mean,
which is in accord with the theoretical result \cite{serota}
\begin{equation}
\langle I \rangle = 0, \quad \Phi = \pm 0.25 h/e
\label{zeromean}
\end{equation}
and which we have also verified by calculating the mean numerically. It is not trivial that the numerical data for $I = \sum_{ \forall E_n \leq E_F} I_n$ plotted in dependence on $E_F$ are centered symmetrically around
zero mean current. We stress that the $I(E_F)$ dependence (the cloud of the data points in the figure \ref{Fig-5}) would become strongly asymmetric when just a single electron level is omitted (mistakenly
or intentionally) from the sum $\sum_{ \forall E_n \leq E_F} I_n$.
It is just this symmetry around zero mean, which allows us to calculate the typical current
by means of a very efficient trick. Now we explain the trick in detail.

\begin{figure}[t]
\centerline{\includegraphics[clip,width=1.0\columnwidth]{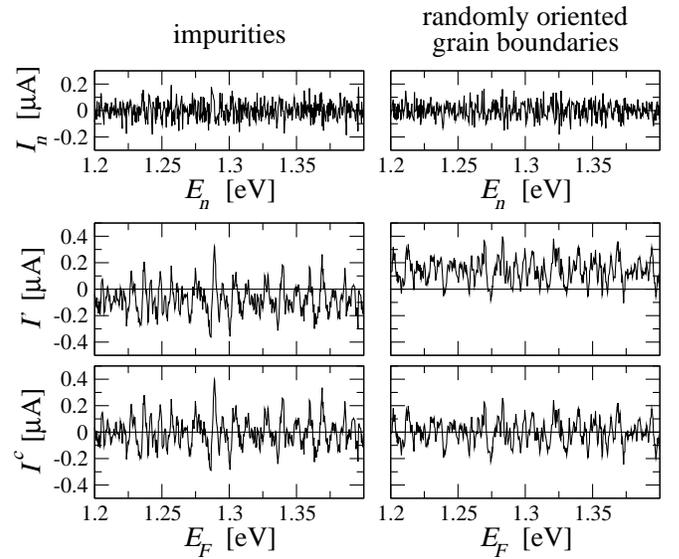}}
\vspace{-0.15cm} \caption{The first row of panels shows the single-electron currents $I_n$ from the preceding figure for the energy window $\delta E = 0.2$eV centered around the energy $E = 1.3$eV. The second row of panels shows the current $I'(E_F) = \sum_{ E_{min} \leq E_n \leq E_F} I_n$, where we sum over the energy levels in the window up to a level $E_n = E_F$.
The third row of panels shows the current $I^{c}(E_F) = I'(E_F) - \langle I' \rangle$, where $\langle I' \rangle$ is the mean obtained by averaging
over all $E_F$ in the window $\delta E$. Unlike $I'(E_F)$, the dependence $I^{c}(E_F)$ is centered around zero mean. Note that the data points are connected by a full line which serves as a guide for eye but obscures the discrete character of the data, seen in the previous figure.} \label{Fig-8}
\end{figure}

The first row of panels in the figure \ref{Fig-8} shows the single-electron currents $I_n$ from figure  \ref{Fig-5} once again, but only for the energy window $\delta E = 0.2$eV centered around the energy $E = 1.3$eV. The second row of panels shows the current $I' = \sum_{ E_{min} \leq E_n \leq E_F} I_n$, where we sum over the energy levels in the window $\delta E$ from the minimum $E_n$ up to $E_n = E_F$. Notice that the data for $I'(E_F)$ are not centered around zero mean.
The third row of panels in the figure \ref{Fig-8} shows the current $I^{c}(E_F) = I'(E_F) - \langle I' \rangle$, where $\langle I' \rangle$ is the mean obtained by averaging
over all $E_F$ in the window $\delta E$. In other words, the dependence $I^{c}(E_F)$ is just the dependence $I'(E_F)$, but centered around zero mean
artificially. Now we are ready to examine the typical persistent current.

\begin{figure}[t]
\centerline{\includegraphics[clip,width=1.0\columnwidth]{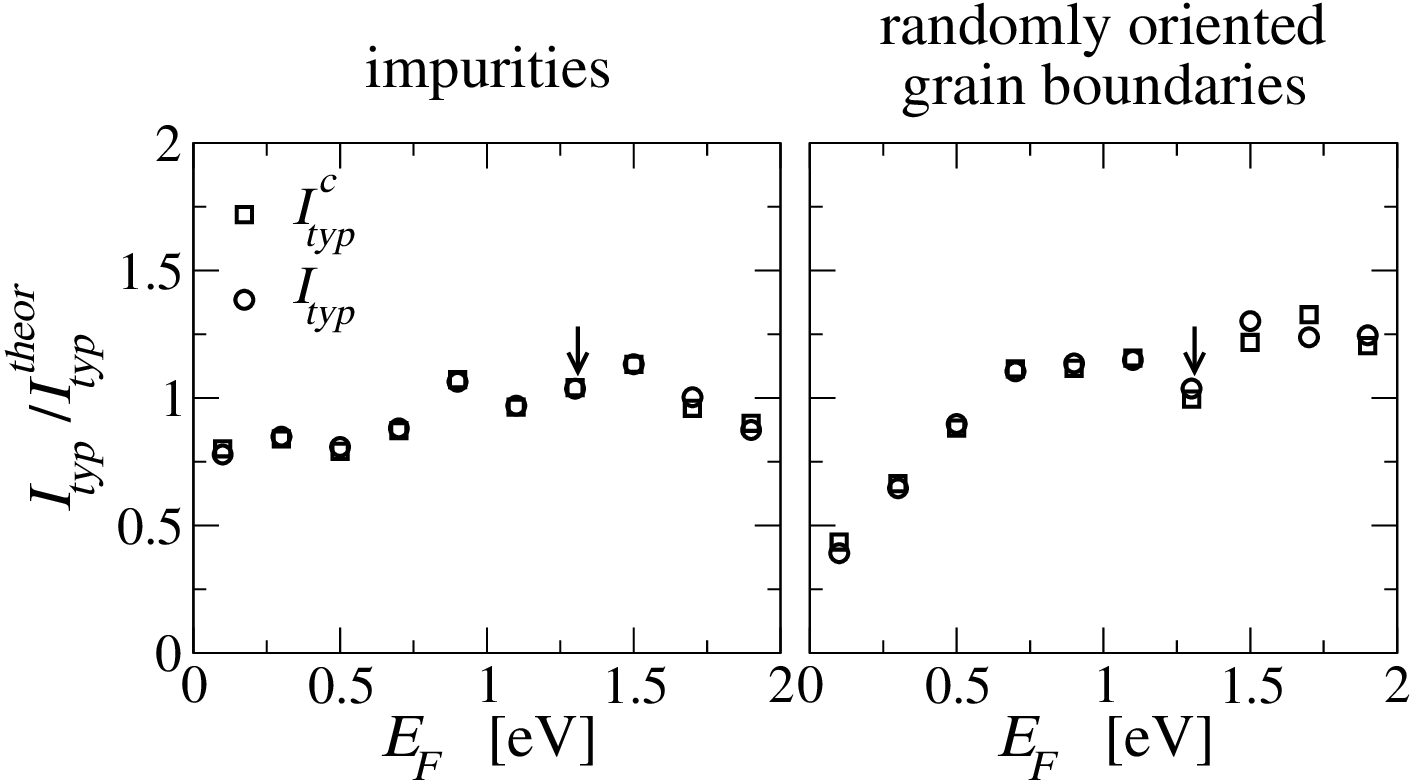}}
\vspace{-0.15cm} \caption{Typical persistent current $I_{typ}$ versus Fermi energy $E_F$, normalized to the theoretical value $I^{theor}_{typ} = 1.6 (e v_F/L) (l/L)$. The circles show
 $I_{typ} = \sqrt{\langle {I}^2 \rangle}$, where $I = \sum_{ \forall E_n \leq E_F} I_n$ is the persistent current due to all single-electron currents below the Fermi level (figure \ref{Fig-5}) and
 $\langle \dots \rangle$ means averaging in the interval $\delta E_F = 0.2$eV around $E_F$. The squares show
 $I^c_{typ} = \sqrt{\langle {I'}^2 \rangle - {\langle {I'} \rangle}^2}$, where $I' = \sum_{ E_{min} \leq E_n \leq E_F} I_n$
 is the current obtained by summing solely the single-electron currents from the energy window $\delta E$ around $E_F$, as discussed in figure  \ref{Fig-8}.
 The values $I^c_{typ}$ originating directly from the data in figure \ref{Fig-8} are labeled by arrows.} \label{Fig-6}
\end{figure}

The figure \ref{Fig-6} shows the typical persistent current $I_{typ}$, calculated in dependence on the Fermi energy $E_F$ and normalized to the theoretical value $I^{theor}_{typ} = 1.6 (e v_F/L) (l/L)$. The open circles show the numerical data for $I_{typ} = \sqrt{\langle {I}^2 \rangle}$, where $I = \sum_{ \forall E_n \leq E_F} I_n$ is the persistent current due to all single-electron currents below the Fermi level.
Such calculation is computationally cost because one has to determine all single-electron eigen-energies $E_n$ below
the Fermi level.

 However, the figure \ref{Fig-6} also shows the numerical data (squares) for the typical current $I^c_{typ} = \sqrt{\langle {I'}^2 \rangle - {\langle {I'} \rangle}^2}$, where $I' = \sum_{ E_{min} \leq E_n \leq E_F} I_n$
 is the current obtained by summing solely the single-electron currents from the energy window $\delta E$ around $E_F$, as discussed in figure  \ref{Fig-8}. This approach works much faster because it is no longer necessary to determine all $E_n$ below
the Fermi level. Indeed, one only needs to determine all $E_n$ in a small energy window $\delta E$ centered around $E_F$. The value
of $\delta E$ should be much larger than the typical inter-level distance, but keeping $\delta E \ll E_F$ still saves a lot of computational time.
The figure \ref{Fig-6} shows that the data for $I^c_{typ}$
reproduce the data for $I_{typ}$ very well.

Moreover, it can be seen that both calculations agree quite well
with the theoretical value $I^{theor}_{typ} = 1.6 (e v_F/L) (l/L)$. The exception are the data in the wire with grain boundaries at small Fermi energies. These data deviate
from $I^{theor}_{typ}$ due to onset of the localization regime at small Fermi energies. (A closer inspection also shows why such deviation is not observed for the impurity disorder. The reason is that the mean free path
decays with the Fermi energy much slower than in the case of the grain-boundary disorder).

\begin{figure}[t]
\centerline{\includegraphics[clip,width=1.0\columnwidth]{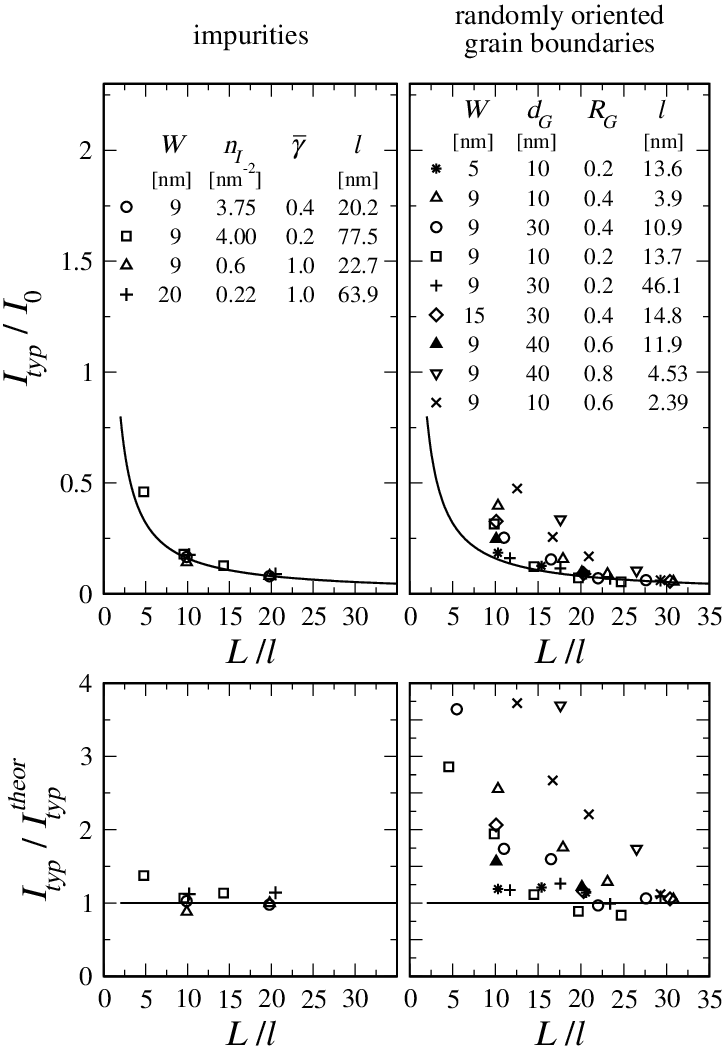}}
\vspace{-0.15cm} \caption{Top panels: Typical persistent current $I_{typ}$ versus the ring length $L$, with $I_{typ}$  normalized to $I_0= ev_F/L$ and with $L$ normalized to the mean free path $l$. The symbols show our numerical results obtained for various parameters listed in the figure. Other parameters are $m=9.109\times10^{-31}$kg and $E_F = 5.6$eV.
The theoretical result $I^{theor}_{typ} = 1.6 (e v_F/L) (l/L)$ is shown in a full line. Bottom panels: The same data as in the top panels, but with $I_{typ}$ normalized to $I^{theor}_{typ}$. } \label{Fig-7a}
\end{figure}

In what follows we speak about the typical current $I_{typ}$ but we in fact evaluate $I^c_{typ}$. We consider the Au rings with material parameters
$m=9.109\times10^{-31}$kg and $E_F = 5.6$eV.
In figure \ref{Fig-7a} the typical persistent currents in rings with impurity disorder and rings with randomly-oriented grain boundaries are studied with impact on the length dependence. The numerical data for $I_{typ}$ (shown by symbols) are obtained for various ring parameters and compared
with the theoretical result $I^{theor}_{typ} = 1.6 (e v_F/L) (l/L)$.
For the impurity disorder the numerical data agree with the formula $I^{theor}_{typ} = 1.6 (e v_F/L) (l/L)$ very well and for the grain boundaries the agreement is also very good for large enough $L/l$. We conclude that for large enough $L/l$ not only the impurity disorder but also the randomly-oriented grain boundaries behave like the white-noise-like disorder.

 However, it can also be seen, that for $L/l$ as large as $\sim 10 - 20$ the typical current in the ring with the randomly-oriented grain boundaries can exceed
 $I^{theor}_{typ}$ by a factor of three to four, when the grain-boundary reflectivity $R_G$ is large and/or the grain size $d_G$ is small. It is remarkable that this happens for the ring lengths for which the corresponding wire resistivity is in the diffusive regime (see the left panel of the figure \ref{scaling}). Of course, the factor of three to four is too small to explain the huge persistent currents ($\sim e v_F/L$) measured \cite{Chand} in a single Au ring of length $L/l \sim 100$. However, it is large enough to resemble the experiment \cite{Jariwala},
 where the measured typical currents exceeded the formula $I^{theor}_{typ} \simeq (ev_F/L)(l/L)$ about two-to-three times.

\subsection{C. Extension to the 3D conductors with columnar grains}

So far we have studied the polycrystalline wires/rings made of the 2D conductor of finite width (figure \ref{Fig:1}).
It is intuitively clear that the obtained results are representative also for the polycrystalline wires/rings made of the 3D conductor,
if the grain boundaries in the conductor are randomly oriented in the 3D space. To extend our numerical study to such 3D systems
is therefore not meaningful.

\begin{figure}[t]
\centerline{\includegraphics[clip,width=0.9\columnwidth]{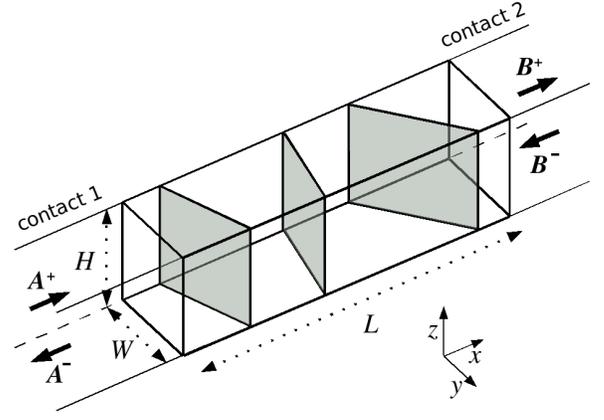}}
\vspace{-0.15cm} \caption{Model of the bamboo-like 3D wire with columnar grains. The columnar grain is a grain shaped as a column
parallel with the growth direction - in our case with the axis $z$. In the bamboo-like wire shown in the figure,
the columnar grains are simply the wire segments separated by the planar boundaries (shaded areas)
randomly oriented with respect to the sidewalls. When viewed from the top, such 3D wire
looks exactly like the 2D wire with the line-shaped grain boundaries (figure \ref{Fig:1}), analyzed up to now.} \label{Fig-14}
\end{figure}

It is however meaningful to extend our study to the 3D wires/rings with columnar grains \cite{Thompson,Thompson2,Thompson3,Liu,Mazor,Faurie,Harris,Thornton,Yeager,Miller}.
In particular, we would like to pay attention (see figure \ref{Fig-14}) to the bamboo-like 3D wires with the columnar grains
separated by the planar boundaries oriented randomly with respect to the wire sidewalls. In reality, the bamboo-like wires \cite{Arzt,Austin,WebbWashburn,Neuner} with the columnar grains
can be viewed as an opposite limit to the polycrystalline wires composed of the tiny 3D grains (with typical size much smaller that the wire cross-section) oriented randomly in the wire volume.  The bamboo-like 3D wires in the figure \ref{Fig-14} are a reasonable idealization
of the real bamboo-like wires, and we will see that the diffusive persistent
currents in the rings made of such wires are remarkably larger that the white-noise-based prediction $I^{theor}_{typ} \simeq (ev_F/L)(l/L)$.

We assume (figure \ref{Fig-14}) that the wire of the width $W$ and thickness $H$ is connected to the semi-infinite contacts. The wave function of the electron at the Fermi level is described by the 3D \schre equation
\begin{equation}
\left[ - \frac{\hbar^2}{2m} \Delta + V_y (y) + V_z (z) + V_D \left( x,y \right) \right] \psi(\vec{r}) = E_F\psi(\vec{r}),
\label{schr3D}
\end{equation}
where $\Delta = \partial^2/\partial x^2 + \partial^2/\partial y^2 + \partial^2/\partial z^2$, $V_D \left( x,y \right)$ is the potential of disorder due to the columnar grain boundaries, and $V_y (y)$ and $V_z (z)$ are the confinement potentials:
\begin{equation}
V_y(y) = \left\{
                          \begin{array}{ll}
                            0, & 0<y<W \\
                            \infty, & \mbox{elsewhere}
                          \end{array} \right.
,
V_z(z) = \left\{
                          \begin{array}{ll}
                            0, & 0<z<H \\
                            \infty, & \mbox{elsewhere}
                          \end{array} \right.
.
\label{steny3D}
\end{equation}

First we solve \eqref{schr3D} in the contacts, where we keep $V_D = 0$ as is customary in the Landauer conductance theory \cite{Datta-kniha}. For $V_D = 0$ the energies in the directions $y$ and $z$,
$\epsilon^W_m$ and $\epsilon^H_n$, are
\begin{equation}
\epsilon^W_m = \frac{\hbar^2 \pi^2}{2mW^2} m^2 , \quad \epsilon^H_n=
\frac{\hbar^2 \pi^2}{2mH^2} n^2, \quad m,n =1,2, \dots ,
\label{eigenenergies y z}
\end{equation}
and the corresponding wave functions are
\begin{equation}
\chi^W_m (y) = \sqrt{\frac{2}{W}} \sin \left( \frac{\pi m}{W}y \right) , \quad \chi^H_n (z) = \sqrt{\frac{2}{H}} \sin \left( \frac{\pi n}{H}z \right) .
\label{eigenfunctions y z}
\end{equation}
The wave function in the contacts can thus be expressed as
\begin{equation}
\psi(\vec{r}) = \sum^{\infty}_{m=1} \sum^{\infty}_{n=1} \left[ a^{+}_{mn} e^{ik_{mn}x} + a^{-}_{mn} e^{-ik_{mn}x} \right] \chi^W_m (y) \chi^H_n (z),
\label{3Droz}
\end{equation}
where the wave vectors $k_{mn}$ obey the equation
\begin{equation}
E_F = \frac{\hbar^2 k^2_{mn}}{2 m} + \epsilon_{mn}, \:\:  \:\: \epsilon_{mn} \equiv \epsilon^W_m + \epsilon^H_n,
\label{Fermivector y z}
\end{equation}
with $\epsilon_{mn}$ being the bottom energy of the channel $[m,n]$.
Clearly, $k_{mn}$ is the Fermi wave vector in the channel $[m,n]$.
The vectors $k_{mn}$ are real for $E_F \ge \epsilon_{mn}$ and imaginary for $E_F < \epsilon_{mn}$. The number of the conducting channels
(channels with $\epsilon_{mn} \le E_F$) is $N_c = \pi N_W N_H / 4$, where $N_W= k_FW/\pi$ and $N_H= k_FH/\pi$ are the numbers of the conducting channels in the $y$ and $z$ directions, respectively. We can order the terms $[mn]$ in the sum \eqref{3Droz} so that the energies
$\epsilon_{mn}$ are ordered increasingly starting by $\epsilon_{11}$. Then the first $N_c$ terms in the sum \eqref{3Droz} are due to the conducting channels.

Now we analyze \eqref{schr3D} in the disordered region, where $V_D (x,y)$ is not zero. Since $V_D (x,y)$ is $z$-independent, it is useful to set into \eqref{schr3D} the expansion $\psi(\vec{r}) = \sum_{n'=1}^{\infty} \varphi_{n'}(x,y)\chi^H_{n'}(z)$. Multiplying \eqref{schr3D} by $\chi^H_{n}(z)^*$ and integrating over $z$ we get the equations
\begin{eqnarray}
\left[ - \frac{\hbar^2}{2m} \left( \frac{\partial^2}{\partial x^2} + \frac{\partial^2}{\partial y^2} \right) + V_y (y) + V_D \left( x,y \right) \right] \varphi_n(x,y) = \nonumber \\
= E_n\varphi_n(x,y), \ \ \ \
\label{schr3Dplacky}
\end{eqnarray}
where $n = 1, 2, \dots$ and $E_n = E_F - \epsilon^H_n$ are the Fermi energies in the 2D subbands arising in the vertical direction.
The equation \eqref{schr3D} thus splits into a set of equations \eqref{schr3Dplacky} which are formally the same as the \schre equation \eqref{schrodgen} for the 2D conductor. Hence, the disordered 3D conductor in figure \ref{Fig-14} can be viewed as a parallel connection of independent 2D conductors with the same disordered potential $V_D(x,y)$, but with various Fermi energies $E_n$. Of course, these 2D conductors are in fact the 2D subbands in the vertical direction.

Therefore, the Landauer conductance of the wire with columnar grains can be expressed as
\begin{equation}
g = \sum\limits_{n=1}^{N_H} g_n(E_n),
 \label{Land3Dto2D}
\end{equation}
where
\begin{equation}
g_n(E_n)
=  \sum\limits_{j=1}^{N_W(E_n)} \sum\limits_{m=1}^{N_W(E_n)} |t_{jm}(E_n)|^2 \frac{k_{j}(E_n)}{k_{m}(E_n)},
 \label{plackaLand3Dto2D}
\end{equation}
is the Landauer conductance \eqref{Land} rewritten for the $n$-th 2D conductor ($n$-th vertical 2D subband) with Fermi energy $E_n$. We recall that $k_{j}(E_n) = k_{jn}$, where $k_{jn}$ is the Fermi wave vector in the 1D channel $[j,n]$, defined by equations \eqref{3Droz} and \eqref{Fermivector y z}. The transmission amplitudes $t_{jm}$, describing the electron transmission through
the columnar grain boundaries, can be evaluated by means of the same scattering matrix as we have introduced in section II for the line-shaped grain boundaries (figure \ref{zrna}), except that now the Fermi energy is $E_n$.

\begin{figure}[t]
\centerline{\includegraphics[clip,width=1.0\columnwidth]{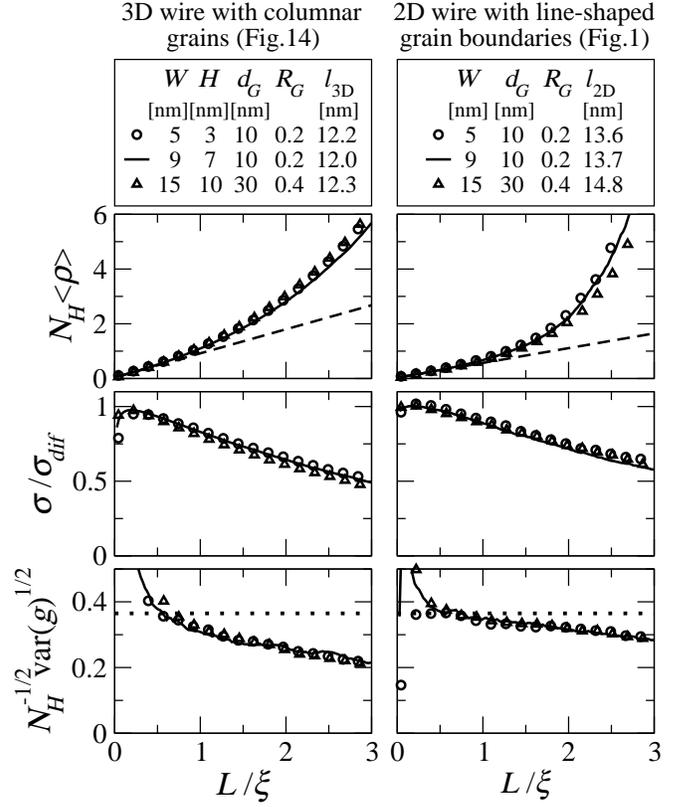}}
\vspace{-0.15cm} \caption{Mean resistance $\langle \rho \rangle$,
conductivity $\sigma$, and conductance fluctuations $\sqrt{\mbox{var}(g)}$ versus $L/\xi$ for the Au wire with grain boundaries.
The left column of panels shows the results for the 3D wire with the columnar grains (figure \ref{Fig-14}),
calculated for three different sets of the parameters $W$, $H$, $d_G$, and $R_G$.
The right column of panels shows the results for the 2D wire with the line-shaped grain boundaries,
which is "fabricated" from the 3D wire with the columnar grains by setting $H \rightarrow 0$ and $N_H = 1$,
and by keeping the same Fermi energy $E_F = 5.6$eV. In other words, the 2D wire and the first vertical
2D subband of the 3D wire are the microscopically-identical 2D conductors. The dashed lines show the linear fit of the $\langle \rho \rangle$ versus $L$ dependence, from which we obtain the diffusive resistivity $\rho_{dif}$. The mean free paths $l_{3D}$ and $l_{2D}$
are extracted from $\rho_{dif}$
and from the 3D and 2D Drude-resistivity expressions.
The conductivity $\sigma$ is obtained from the mean conductance $\langle g \rangle$ by means of \eqref{sigmag} and normalized by $\sigma_{dif}=1/\rho_{dif}$.} \label{Fig-15}
\end{figure}

In the figure \ref{Fig-15}, transport in the 3D wire with columnar grains is compared with transport in the
corresponding 2D wire,
obtained from the 3D wire by setting $H \rightarrow 0$ and $N_H = 1$,
and by keeping the same Fermi energy. This means that the grain boundaries in the 2D wire are the randomly-oriented line-shaped boundaries studied in the preceding text (right sketch in the figure \ref{Fig:1}).
The comparative study shows a few results which are worth to stress.

First a comment on the localization length $\xi$ in the figure \ref{Fig-15}. Note that the values of $\xi$ in the 3D wire and 2D wire
are the same. The 3D wire with the columnar grains is a parallel connection of $N_H$
independent 2D wires (2D subbands) with the same disorder and different Fermi energies. The 2D wire with the largest
Fermi energy provides the largest localization length and this is just the localization length of the whole
3D wire because the conductance contributions from other $N_H - 1$ wires become negligible for large $L$. The 2D wire with the largest Fermi energy is just the 2D wire
obtained from the 3D wire by setting $H \rightarrow 0$
and by keeping the same Fermi energy. As a result, $\xi$ is the same in the 3D and 2D wires.

Notice now the mean resistance for $L < \xi$. It is roughly $N_H$ times smaller
for the 3D wire than for the 2D wire and the resulting mean-free paths $l_{3D}$ and $l_{2D}$
give the numerical ratio $l_{3D}/l_{2D} \simeq 0.9$, which is in good accord with the
formula
\begin{equation}
l_{3D} = \frac{9 \pi}{32} l_{2D} \simeq 0.88 l_{2D},
\label{l3Dto2D maintext}
\end{equation}
derived in the appendix B. One also sees that for $L < \xi$ the conductivity $\sigma$
exhibits in the 3D wire essentially the same weak localization behavior as in the 2D wire.
In summary, \emph{the resistance and conductance of the 3D wire with columnar grains
exhibit a standard diffusive behavior, similarly as for disorder which is white-noise-like in the 3D space.}
The fact that the wire is effectively composed of the $N_H$ independent 2D wires is reflected
by the conductance fluctuations: the figure \ref{Fig-15} shows that $\sqrt{\mbox{var}(g)}$ is roughly $\sqrt{N_H}$ times larger than
the standard value, which one expects.

Consider now the ring made of the 3D wire with the columnar grains.
The ring is composed of the $N_H$ independent 2D rings. If the $n$-th 2D ring carries the persistent
current $\mathcal{I}_{n}$, the total persistent current $I_{CG}$ in the ring with the columnar grains reads
\begin{equation}
I_{CG} = \sum_{n=1}^{N_H} \mathcal{I}_{n}(E_n).
\label{I_CG_singlering}
\end{equation}
To calculate $\mathcal{I}_{n}$ numerically, we evaluate for each individual 2D ring
the spectrum of all single-electron currents below the Fermi level (in the same way as in the figure \ref{Fig-5}) and
we sum these currents to obtain $\mathcal{I}_{n}$. After that we evaluate the sum \eqref{I_CG_singlering}
and we eventually perform averaging to obtain the typical current $I^{CG}_{typ} = \sqrt{\langle I_{CG}^2 \rangle}$.

\begin{figure}[t]
\centerline{\includegraphics[clip,width=1.0\columnwidth]{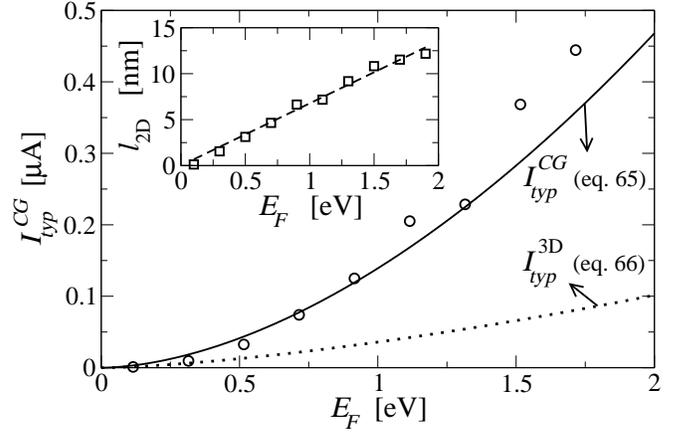}}
\vspace{-0.15cm} \caption{Typical persistent current in the 3D ring with the columnar grains, $I^{CG}_{typ}$,
versus the Fermi energy. The ring dimensions are $W=9nm$, $H=7nm$, and $L = 130$nm ($L \gg l$ for all considered $E_F$), the parameters
of the grain boundaries are $R_G(E_F = 2eV) = 0.2$ and $d_G = 10nm$.
The circles show the numerical data, the full line represents the estimate \eqref{ItypCGto3D text},
and the dotted line is the white-noise-based result \eqref{i3Digre text}.
Inset shows the mean free path in the 2D wire, $l_{2D}$, as a function of the Fermi energy:
the squares show the numerical data and the dotted line is the linear fit $l_{2D} = C E_F$.} \label{Fig-166}
\end{figure}

We also estimate $I^{CG}_{typ}$ analytically. The simplest estimate,
$I^{CG}_{typ} \simeq \sqrt{N_H} (ev_F/L)(l_{2D}/L)$, assumes that each of the $N_H$ rings
supports the same typical current (the value of which is $(ev_F/L)(l_{2D}/L)$ because
the columnar grains create the white-noise-like 2D disorder in the plane perpendicular to the columns).
A more precise estimate (appendix C) gives
\begin{equation}
I^{CG}_{typ} = \frac{64}{3 \sqrt{35} \pi} \sqrt{N_H} I^{3D}_{typ} \simeq 1.26 \sqrt{N_H} (ev_F/L)(l_{3D}/L),
\label{ItypCGto3D text}
\end{equation}
where 
\begin{equation}
I^{3D}_{typ} = 1.1 (ev_F/L)(l_{3D}/L)
\label{i3Digre text}
\end{equation}
is the expression \eqref{Igre} for $d = 3$, the 3D mean free path $l_{3D} = 0.88 l_{2D}$ (equation \ref{l3Dto2D maintext}),
and the 2D mean free path is assumed in the form $l_{2D} = C E_F$ (the constant $C$ is determined by fitting the numerically calculated $l_{2D}$ as shows the inset to the figure \ref{Fig-166}).

In the figure \ref{Fig-166} the typical current in the ring with the columnar grains is calculated as a function of the Fermi energy.
It can be seen that the formula
\eqref{ItypCGto3D text} (full line) agrees quite well with the numerically calculated $I^{CG}_{typ}$ (open circles), while
 the formula \eqref{i3Digre text} (dotted line) underestimates the numerical data about $\sqrt{N_H}$ times. Obviously,
 the formula \eqref{i3Digre text} holds only for disorder which is white-noise-like in the 3D space, not the case for the columnar grains.

 The ring considered in the figure
 \ref{Fig-166} is rather small. For the Au ring with $H = 70$nm the formula \eqref{ItypCGto3D text} gives the result
 $I_{typ} \simeq 20 (ev_F/L)(l_{3D}/L)$. This result resembles the experiment \cite{Chand}, where
 the diffusive persistent currents in three individual Au rings of length $L/l \sim 100$
 exceeded the value $(ev_F/L)(l_{3D}/L)$ one-to-two order of magnitude.

\section{V. Summary and concluding remarks}

\subsection{A. Summary of results}

We have studied mesoscopic transport in the Q1D wires and rings
made of a 2D conductor of width $W$ and length $L \gg W$. We have compared transport in an impurity-free conductor with grain boundaries with transport in a grain-free conductor with impurity disorder.

The transmission through the disordered conductors was calculated by
the scattering-matrix method, and the Landauer conductance has been obtained.
We have also calculated the persistent current in the rings threaded by magnetic flux: we have incorporated into the scattering-matrix method the
flux-dependent cyclic boundary conditions and we have introduced a trick allowing to study the persistent currents in rings of almost realistic size (the typical persistent current for magnetic flux $\pm 0.25 h/e$
 was extracted from the single-electron energies in a narrow window around the Fermi energy). We have mainly studied the conductance
 and persistent current in the diffusive transport regime. Our results are the following.

 If the grain boundaries are weakly reflecting, the systems with the randomly-oriented grain boundaries show the same (mean) conductance and the same (typical) persistent current as the systems with impurities. The obtained results also agree with the single-particle theories of diffusive transport \cite{Datta-kniha,Cheung,riedel}, treating disorder as a white-noise-like potential.

However, if the grain boundaries are strongly reflecting, the rings with the randomly-oriented grain boundaries
  can exhibit in the diffusive regime the typical persistent currents about three-to-four times larger than the white-noise-based formula $I^{theor}_{typ} \simeq (ev_F/L)(l/L)$. This finding resembles the experiment \cite{Jariwala},
   where the typical persistent currents measured in the diffusive Au rings were two-to-three times larger than $I^{theor}_{typ} \simeq (ev_F/L)(l/L)$.

We have also extended our study to the 3D conductors with the columnar grains.
We have shown that the typical persistent current in the diffusive metallic ring with the columnar grains is given by the formula
$I_{typ} \simeq 1.26 \sqrt{N_H} (ev_F/L)(l/L)$, where $N_H \simeq Hk_F/\pi$ is the number of the 2D subbands within the thickness $H$.
For the Au ring with $H = 70$nm  the formula gives the result
 $I_{typ} \simeq 20 (ev_F/L)(l/L)$, which is not far from the experiment \cite{Chand}, where
 the diffusive persistent currents measured in three individual Au rings of length $L/l \sim 100$
 were one-to-two orders of magnitude larger than $I^{theor}_{typ} \simeq (ev_F/L)(l/L)$.

 \subsection{B. Comment on relevance for experiment}

 Of course, we cannot conclude that our study is a definite explanation of the experiments \cite{Chand} and \cite{Jariwala}, because
 the polycrystalline structure of the experimental samples in these works is not known. Moreover, even if we would assume that the polycrystalline grains
 in the experiment \cite{Chand} are both columnar and strongly reflecting, our study predicts
 the persistent current only thirty-to-fifty times larger than the formula $I^{theor}_{typ} \simeq (ev_F/L)(l/L)$ while the largest experimental value
 in \cite{Chand} exceeds the value $(ev_F/L)(l/L)$ almost two hundred times.

 Nevertheless, according to our study one should not be surprised when two experiments  \cite{Bluhm,Bles} confirm the formula $I^{theor}_{typ} \simeq (ev_F/L)(l/L)$ convincingly and two other experiments \cite{Chand,Jariwala} do not. Our study shows clearly within the single-particle picture, that the experimental results can depend
 quite strongly on the nature of the polycrystalline grains, being very different for different fabrication conditions \cite{Thompson,Thompson2,Thompson3,Liu,Mazor,Faurie,Harris,Thornton,Yeager,Miller}.

 The columnar grains are
fundamentally different from the tiny randomly-oriented grains, implicitly assumed in any 3D white-noise-based description of disorder.
Unlike the tiny random grains, the columnar grains produce the white-noise-like disorder only in the plane perpendicular to the columns,
which gives rise to the factor $\sqrt{N_H}$ in the formula $I_{typ} \simeq 1.26 \sqrt{N_H} (ev_F/L)(l/L)$ but which has essentially
no effect on the diffusive resistance of the wire. It might be instructive to fabricate intentionally the diffusive normal-metal rings with various types
of grains, to measure the persistent current, and to correlate the data with the grain properties: the results $I_{typ} \simeq (ev_F/L)(l/L)$ and $I_{typ} \simeq \sqrt{N_H} (ev_F/L)(l/L)$ should appear for the tiny random grains
and columnar grains, respectively.

\subsection{C. Comment on large persistent currents in rings with perpendicular grain boundaries }

Finally, we make a comment on the rings and wires with the grain boundaries perpendicular to the current. By considering the perpendicular boundaries, the work \cite{Kircze} predicted the persistent currents of size $\sim e v_F/L$ in rings of length $L \gg l_{\langle g \rangle}$, where $l_{\langle g \rangle}$ is the 'mean free path' defined as
  \begin{equation}
   l_{\langle g \rangle} = \frac{2L}{k_FW}\langle g \rangle,
   \label{lkirczenov}
   \end{equation}
   with $\langle g \rangle$ being the corresponding wire conductance. Since the result $\sim e v_F/L$ strongly resembles the experimental results \cite{Chand},
   we revisit it briefly. We apply our 2D model (figure \ref{zrna}).

\begin{figure}[t]
\centerline{\includegraphics[clip,width=1.0\columnwidth]{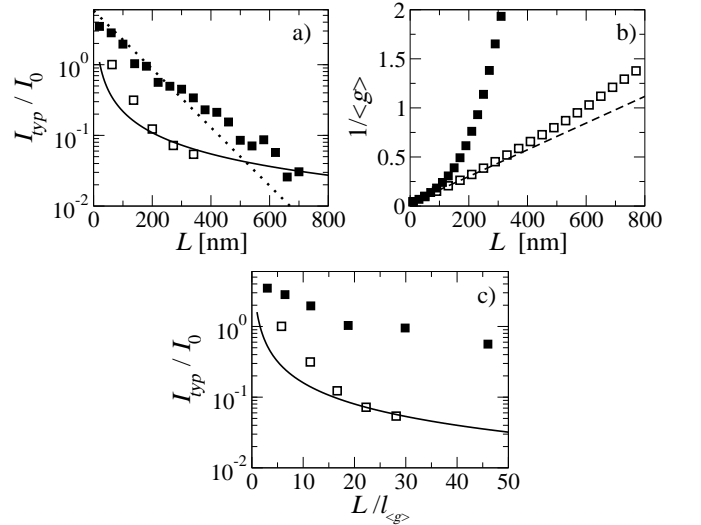}}
\vspace{-0.15cm} \caption{Panel \emph{a}: Typical persistent current $I_{typ}$ versus the ring length $L$, normalized to $I_0= ev_F/L$. The empty squares are the numerical data for the randomly-oriented grain boundaries.
The full squares are the numerical data for the perpendicular grain boundaries.
The full line shows the formula $I^{theor}_{typ} = 1.6 (e v_F/L) (l/L)$. The dotted line shows the formula  \eqref{localicur}, where $\xi = 3.7 l$ is the localization length determined earlier. The mean free path $l$ in these formulae originates from the wire resistivity $\rho_{dif} = 2/k_F l$, determined in subsection IV.A. Panel \emph{b}: Inverse mean conductance $1/\langle g \rangle$ versus the wire length $L$ in the wires corresponding to the rings in the panel \emph{a}. The dashed line is the diffusive dependence  $1/\langle g \rangle = 1/N_c + (2/k_F l)(L/W$). Panel \emph{c}: The same numerical data as
in the panel \emph{a}, but versus the ratio $L/l_{\langle g \rangle}$, where $l_{\langle g \rangle}$ is extracted from the data in panel \emph{b} by means of the formula  \eqref{lkirczenov}
.}
\label{irad}
\end{figure}

  In the figure \ref{irad}\emph{a} we compare the typical persistent currents in the rings with the randomly oriented boundaries (empty squares)
  and rings with the perpendicular boundaries  (full squares), while in the  figure \ref{irad}\emph{b} we show the corresponding wire conductances
  $\langle g \rangle$. Note that these data are plotted in dependence on the length $L$. We set the conductances from figure \ref{irad}\emph{b}
   into the formula \eqref{lkirczenov} and we evaluate $ l_{\langle g \rangle}$ and $L/l_{\langle g \rangle}$. In the figure  \ref{irad}\emph{c}
  we plot the typical currents from figure \ref{irad}\emph{a} in dependence on $L/l_{\langle g \rangle}$.
  Indeed, for the perpendicular boundaries we observe the effect predicted in \cite{Kircze},
  namely the persistent currents $\sim e v_F/L$ for the ring lengths $L/l_{\langle g \rangle} \gg 1$.
However, in reference \cite{Kircze} this result was not compared with the result for the randomly oriented boundaries. The figure \ref{irad}\emph{c}
  shows that the large current diminishes when the orientation of the boundaries becomes random.

  Since the values $\sim e v_F/L$ resemble the large persistent currents in the Au rings of the experiment \cite{Chand}, one might
  speculate about presence of the perpendicular grain boundaries in the measured rings. However, we see in the figure \ref{irad}\emph{b}
  that the wire with the perpendicular boundaries is in the localization regime for all wire lengths $L$, for which we observe the currents $\sim e v_F/L$ in the ring. In contrast to this, the Au wires used to determine the mean free path experimentally \cite{Chand} were safely in the diffusive regime, not in the localization regime.

  Moreover, the formula \eqref{lkirczenov} should not be used when the wire is in the localization regime.
  Indeed, the mean free path should be length independent, while $l_{\langle g \rangle}$ depends on $L$ quite strongly
  due to the exponentially raising  $1/\langle g \rangle$.

  The figure \ref{irad}\emph{a} also shows that the numerical data for the perpendicular boundaries
  roughly agree with the formula
  \begin{equation}
  I_{typ} = \sqrt{N_c} (e v_F/L) \exp(-L/2\xi)
  \label{localicur}
  \end{equation}
  while the numerical data for the randomly oriented boundaries
  approach the formula $I^{theor}_{typ} = 1.6 (e v_F/L) (l/L)$. The formula \eqref{localicur} describes the Q1D ring with $N_c$ non-communicating channels in the localization regime \cite{Cheung1D}, which is not the case for the Q1D rings of the experiment \cite{Chand}.

\section{Acknowledgement}
We thank for the grant VVCE-0058-07, VEGA grant 2/0633/09, and grant APVV-51-003505.

\section{Appendix A: Typical persistent current for magnetic flux $\Phi = \pm 0.25\Phi_0$}

The persistent current $I$ in a single mesoscopic Q1D ring is a periodic function of magnetic flux $\Phi$. The period of the function is $\Phi_0 = h/e$. Therefore, it can be expanded into the Fourier series as
\begin{equation}
I(\Phi) = \sum^\infty_{p=1} I_p \sin(2 \pi p \Phi / \Phi_0).
\end{equation}
The current $I$ in a disordered ring strongly fluctuates from sample to sample due to the microscopic fluctuations of disorder. We are therefore interested in the typical persistent current $I_{typ} = \sqrt{\langle I^2 \rangle}$, where $\langle \dots \rangle$ means averaging over different configurations of disorder.
Assuming the white-noise-like disorder (see the text), the authors of the work \cite{riedel} derived the equation
\begin{equation}
\langle I^2(\Phi) \rangle  = \sum^\infty_{p=1} \langle I^2_p \rangle \sin^2 (2 \pi p \Phi/\Phi_0),
\label{I2odfi}
\end{equation}
where
\begin{equation}
\langle I^2_p \rangle  = \frac{96}{\pi^2 p^3} \left( \frac{e}{\tau_D} \right)^2
\end{equation}
is the mean square of the $p$-th harmonics,
$\tau_D  = L^2/D$
is the electron diffusion time around the ring, and $D = v_F l /d$ is the electron diffusion coefficient.
For $\Phi = \pm 0.25 \Phi_0$ the expression \eqref{I2odfi} can be rewritten as
\begin{equation}
\langle I^2 \rangle  = \frac{96}{\pi^2} \left( \frac{e}{\tau_D} \right)^2 \sum\limits^{\infty}_{p=0} \frac{1}{(2p+1)^3}.
\label{smk}
\end{equation}
We perform summation in \eqref{smk} and we obtain the typical persistent current in the form
\begin{equation}
I^{theor}_{typ} = \sqrt{\langle I^2 \rangle}  \simeq 3.2 \frac{e}{\tau_D}.
\label{Ithtau}
\end{equation}
If we set into \eqref{Ithtau} the above mentioned expressions for $\tau_D$ and $D$, we obtain the equation \eqref{Igre}.

\section{Appendix B: Mean free path in 3D wire with columnar grains}

We set into the equation \eqref{Land3Dto2D} the formulae $g = \frac{WH}{L} \sigma_{3D}$ and $g_n = \frac{W}{L} \sigma^{2D}_n$,
where $\sigma_{3D}$ is the 3D conductivity and $\sigma^{2D}_n$ is the conductivity
of the $n$-th 2D wire. We obtain the equation
\begin{equation}
\sigma_{3D}  = \frac{1}{H} \sum_{n=1}^{N_H} \sigma^{2D}_n .
\label{sigma 3D}
\end{equation}
We set into \eqref{sigma 3D} the Drude expressions
\begin{equation}
\sigma_{3D} = \frac{  k^2_F l_{3D}}{3 \pi }, \quad \sigma^{2D}_n = \frac{k_n l_n}{2},
\end{equation}
where $k_F$ and $l_{3D}$ are the 3D Fermi wave vector and 3D mean free path, and
$k_n = \sqrt{2 m E_n}/\hbar$ and $l_n$ are the Fermi wave vector and mean free path in the $n$-th 2D wire.
We find that
\begin{equation}
l_{3D} = \frac{3}{k_F N_H}\sum^{N_H}_{n=1} \frac{k_n l_{n}}{2}.
\label{NHi1}
\end{equation}
For the 2D wire with the grain boundaries we expect the dependence $l_n \propto E_n$. This dependence is in good accord
with our numerical calculations (see the inset to the figure \ref{Fig-166}).
We therefore set into \eqref{NHi1} the formula $l_n = C E_n$, where $C$ is a constant.
Moreover, we also set into \eqref{NHi1} the expression $E_n/E_F = 1 - n^2/N_H^2$. We obtain the equation
\begin{equation}
l_{3D} =  \frac{3 l_{2D}}{2 N_H} \sum^{N_H}_{n=1} \left( 1 - \frac{n^2}{N_H^2} \right)^{3/2},
\label{NHi13}
\end{equation}
where we have used $l_{n=1} = C E_1 \simeq C E_F \equiv l_{2D}$ (here $l_{2D}$
is the 2D mean free path in the 2D wire with the same $E_F$ as in the 3D wire).
For $N_H \gg 1$ the sum in \eqref{NHi13} can be replaced by integral and we obtain the result
\begin{equation}
l_{3D} = \frac{9 \pi}{32} l_{2D} \simeq 0.88 l_{2D},
\label{l3Dto2D}
\end{equation}
which is in good accord with our simulation (figure \ref{Fig-15}).

\section{Appendix C: Typical persistent current in 3D ring with columnar grains}

Using the formula \eqref{I_CG_singlering}, the typical persistent current $I^{CG}_{typ}$ in the 3D ring with columnar grains can be written as
\begin{equation}
I^{CG}_{typ} = \sqrt{\langle I_{CG}^2 \rangle} = \sqrt{\sum_{n=1}^{N_H} \langle \mathcal{I}^2_{n} \rangle},
\label{persCGtyp}
\end{equation}
where we have utilized the fact that the persistent currents $\mathcal{I}_n$ in the constituting 2D rings are mutually
uncorrelated. Due to the columnar grains, each 2D ring is subjected to the white-noise-like 2D disorder and therefore
carries the typical current given by the formula  \eqref{Igre} with $d = 2$. Thus
\begin{equation}
\langle \mathcal{I}^2_{n} \rangle = \left[ 1.6 \frac{e v_n}{L} \frac{l_n}{L} \right]^2.
\end{equation}
We set the last equation into \eqref{persCGtyp} and we also apply the equation
$v^2_n = 2E_n/m$ and equation $l_n = C E_n$ from the preceding appendix. We obtain the equation
\begin{equation}
I^{CG}_{typ} = 1.6 \frac{e}{L^2} C \sqrt{\frac{2}{m}} \sqrt{\sum_{n=1}^{N_H} E^3_n},
\end{equation}
which can be rewritten into the form
\begin{equation}
I^{CG}_{typ} = 1.6 \frac{e v_F}{L} \frac{l_{2D}}{L} \sqrt{\sum_{n=1}^{N_H} \left(1 - \frac{n^2}{N_H^2} \right)^3 }
\label{i3D}
\end{equation}
by using the same procedure as in the preceding appendix. For $N_H \gg 1$ the sum in \eqref{i3D} can be replaced by integral and calculated analytically. We obtain the formula
\begin{equation}
I^{CG}_{typ} = \frac{4}{\sqrt{35}} \sqrt{N_H} 1.6 \frac{e v_F}{L} \frac{l_{2D}}{L}.
\label{Ityp3D}
\end{equation}
which relates the typical current in the 3D ring with the columnar grains to the typical current
in the 2D ring with the same Fermi energy and disorder. By means of the formula \eqref{l3Dto2D}
one can rewrite \eqref{Ityp3D} into the form \eqref{ItypCGto3D text}.

%

\end{document}